\documentclass{aa}
\usepackage{graphicx}
\usepackage{natbib}

\bibpunct{(}{)}{;}{a}{}{,}

\newcommand{\GILDAS}{\texttt{GILDAS}}
\newcommand{\CLASS}{\texttt{CLASS}}
\newcommand{\CLIC}{\texttt{CLIC}}
\newcommand{\MAPPING}{\texttt{MAPPING}}

\newcommand{\IRAMthm}{\textrm{IRAM-30m}}
\newcommand{\PdBI}{\textrm{PdBI}}

\newcommand{\ie} {{\em i.e.}}
\newcommand{\eg} {{\em e.g.}}

\newcommand{\about}{\emm{\sim}}

\newcommand{\thCO}  {\mbox{$^{13}$CO}}       
\newcommand{\twCO}  {\mbox{$^{12}$CO}}       

\newcommand{\Jone}{\mbox{$J$=1--0}}
\newcommand{\Jtwo}{\mbox{$J$=2--1}}

\newcommand{\emm}[1]{\ensuremath{#1}}   
\newcommand{\emr}[1]{\emm{\mathrm{#1}}} 
\newcommand{\unit}[1]{\emm{\, \emr{#1}}}

\newcommand{\K}   {\unit{K}}
\newcommand{\mm}  {\unit{mm}}
\newcommand{\m}   {\unit{m}}
\newcommand{\kms} {\unit{km\,s^{-1}}}
\newcommand{\Kkms}{\unit{K\,km\,s^{-1}}}
\newcommand{\MHz} {\unit{MHz}}

\newcommand{\Kpccm}{\unit{K\,cm^{-3}}}
\newcommand{\pccm}{\unit{cm^{-3}}}
\newcommand{\pscm}{\unit{cm^{-2}}}
\renewcommand{\mag} {\unit{mag}}

\newcommand{\Av}{\emm{\emr{A_v}}} 

\newcommand{\Tas}{\emm{T_\emr{A}^*}}
\newcommand{\Tmb}{\emm{T_\emr{mb}}}
\newcommand{\Tsys}{\emm{T_\emr{sys}}}
\newcommand{\Beff}{\emm{B_\emr{eff}}}
\newcommand{\Feff}{\emm{F_\emr{eff}}}

\def\AV{\mbox{A$_{\rm V}$}}

\def\HH{\mbox{H$_2$}}

\def\nH2{{\rm n}({\rm H}_2)}
\def\NH2{{\rm N}({\rm H}_2)}
\def\pccc{~{\rm cm}^{-3}} 
\def\pcc {~{\rm cm}^{-2}}

\def\Tstar#1 {\mbox{${\rm T}_{\rm #1}^*$}}
\def\Tsub#1 {\mbox{${\rm T}_{\rm #1}$}}
\def\TK  {\Tsub K }
\def\TB  {\Tsub B }

\def\TR {\Tsub R }

\def\Tmb {\Tsub mb }

\def\arcsec{\mbox{$^{\prime\prime}$}} 
\def\arcmin{\mbox{$^{\prime}$}}
\def\ibid{{\it ibid}}

\def\degr{$^{\rm o}$}
\def\p{\mbox{$^+$}}
\def\twCO {\mbox{$^{12}$CO}}
\def\thCO {\mbox{$^{13}$CO}}

\def\hcop{\mbox{{HCO\p}}}

\def\cch{\mbox{C$_2$H}}
  
\def\hhco{\mbox{H$_2$CO}}
\def\h13cop{\mbox{{H$^{13}$CO\p}}}

\def\c3h2{\mbox{C$_3$H$_2$}}
 
 \def\R0{R$_0$} 
  \def\kpc{\rm kpc}
 \def\deg{{}^\circ}
\def\ddeg{{}^\circ\kern-.1em}  
 \def\pc{\rm pc}

\def\E#1 {$10^{#1}$}
\def\E#1 {E{#1}}
\def\P#1,{$\nH2\TK~=~#1\times~10^4\pccc$~K}
\def\ec#1,#2,#3,{#1\,(#2)\E{#3}}

\def\zoph{$\zeta$ Oph}

\def\H3{\mbox{H$_3$}}

\def\ammon{\mbox{N\H3} }

\def\RH2{\mbox {R$_{\rm G}$}}
\def\fH2{\mbox {f$_{\HH}$}}
\def\FH2{\mbox {F$_{\HH}$}}

\newcommand{\TabObs}{%
  \begin{table*}
    \caption{Observation parameters. The projection center of all the data 
      matches the position of NRAO150: $\alpha_{2000} = 03^h59^m29.747^s$, 
      $\delta_{2000} = 50\deg57'50.16''$.}
    \begin{center}
      {\tiny
        \begin{tabular}{rcrllccccccr}
          \hline \hline
          Molecule & Transition & Frequency  & Instrument & Config. & Beam   & PA     & Vel. Resol. & Int. Time & \Tsys{} & Noise$^{a}$ & \multicolumn{1}{c}{Obs. date} \\
                   &            & GHz        &            &         & arcsec & $\deg$ & \kms{}      & hours     & K       & K           & \\
          \hline
          \twCO{} & \Jone{} & 115.271202 & PdBI & D & $ 6.0 \times 5.5$ & 82 & 0.2 & 3.3/12.5 & 230 & 0.20 & Jul. 2005 \\
          \twCO{} & \Jone{} & 115.271202 & PdBI & C & $ 2.7 \times 2.5$ & 12 & 0.2 & 7.0/12.0 & 230 & 0.52 & Dec. 2005 \\
          \hline
        \end{tabular}}
    \end{center}
    $^{a}$ The noise values quoted here are the noises at the mosaic phase center
    (Mosaic noise is inhomogeneous due to primary beam correction; it 
    steeply increases at the mosaic edges).
    \begin{center}
      {\tiny
        \begin{tabular}{rcrlcccrclccr}
          \hline \hline
          Molecule & Transition & Frequency  & Instrument & \# Pix. & \Feff{} & \Beff{} & Resol. & Resol. & Int. Time$^{a}$ & \Tsys{} & Noise$^{b}$ & Obs. date \\
                   &            & GHz        &            &         &         &         & arcsec & \kms{} & hours           &    K    &  K          &   \\
          \hline
          \twCO{} & \Jone{} & 115.271202 & 30m/AB100 & 2 & 0.95 & 0.75 & 22.5 & 0.20 & 2.4/3.6  & 240 & 0.26 & Sep. 2005 \\
          \twCO{} & \Jtwo{} & 230.538000 & 30m/AB230 & 2 & 0.91 & 0.52 & 22.5 & 0.20 & 2.4/3.6  & 320 & 0.16 & Sep. 2005 \\
          \hline
        \end{tabular}}
    \end{center}
    $^{a}$ Two values are given for the integration time: the on-source
    time and the telescope time.\\
    $^{b}$ Noise values estimated at the position of NRAO150.
    \label{tab:obs}
  \end{table*}}

\newcommand{\TabFlux}{%
  \begin{table}
    \centering
    \caption{NRAO150 flux in Jy.}
    \begin{tabular}{ccc}
      \hline
      \hline
      Date & 2.6\mm{} & 1.3\mm{} \\
      \hline
      17.07.2005 &  2.8 & 1.6 \\
      28.07.2005 &  2.8 & 1.6 \\
      01.08.2005 &  2.9 & 1.7 \\
      12.12.2005 &  2.4 & 1.4 \\
      \hline
    \end{tabular}
    \label{tab:fluxes}
  \end{table}}

\newcommand{\FigIntro}{%
  \begin{figure}
    \includegraphics[width=0.94\hsize{}]{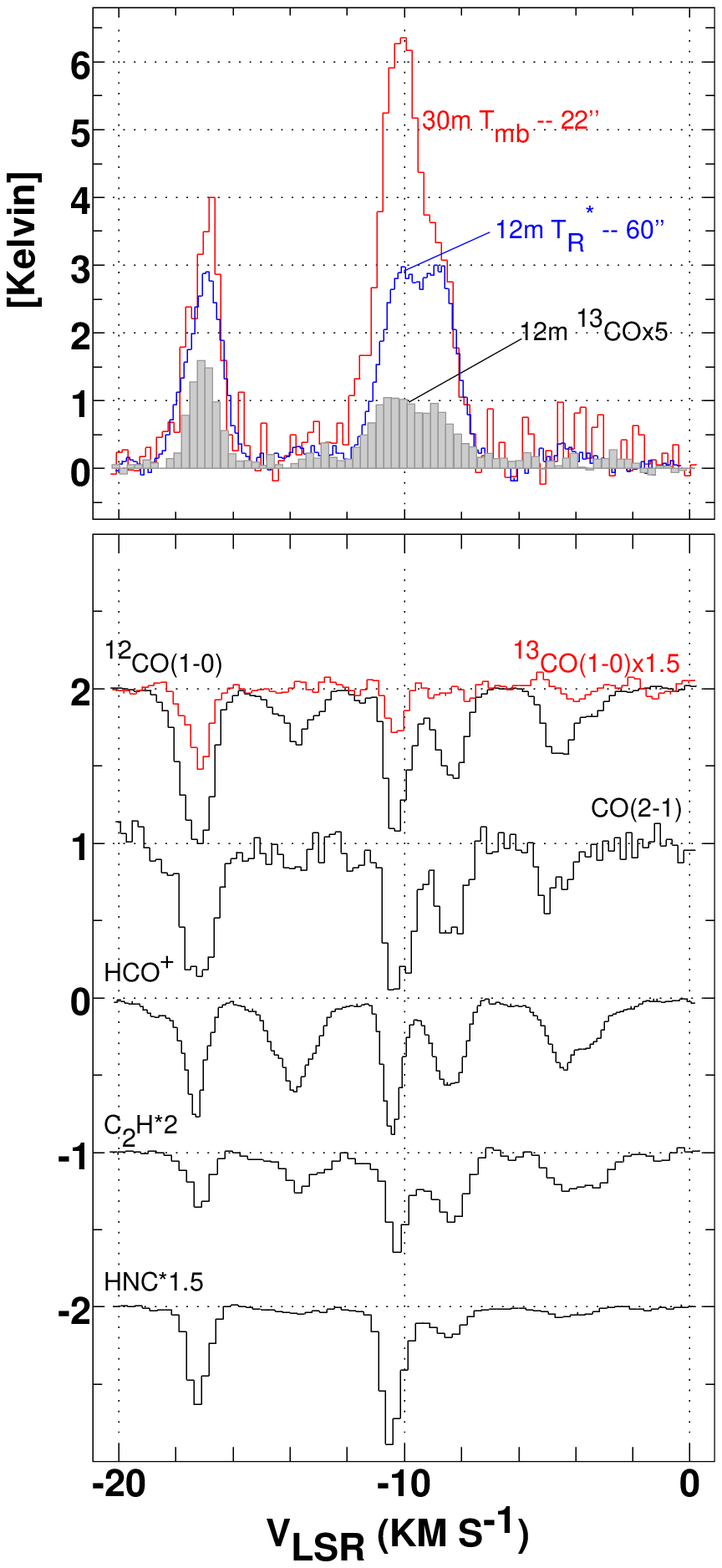}
    \caption{Emission and absorption spectra toward B0355+508 $aka$ NRAO150.  
      {\bf Top panel:} 1\arcmin\ HPBW J=1-0 \twCO\ and \thCO\ profiles from
      the former NRAO 12m telescope on a \TR$^*$ brightness scale and the
      22.5\arcsec-resolution IRAM 30m \twCO\ profile from the present work
      on a main-beam brightness temperature scale.  {\bf Bottom:} IRAM PdBI
      absorption profiles from various representative species.  The
      profiles have been shifted vertically and scaled in some instances.}
    \label{fig:B0355f1}
  \end{figure}}

\newcommand{\FigProfileEvolution}{%
  \begin{figure}
    \includegraphics[width=0.94\hsize{}]{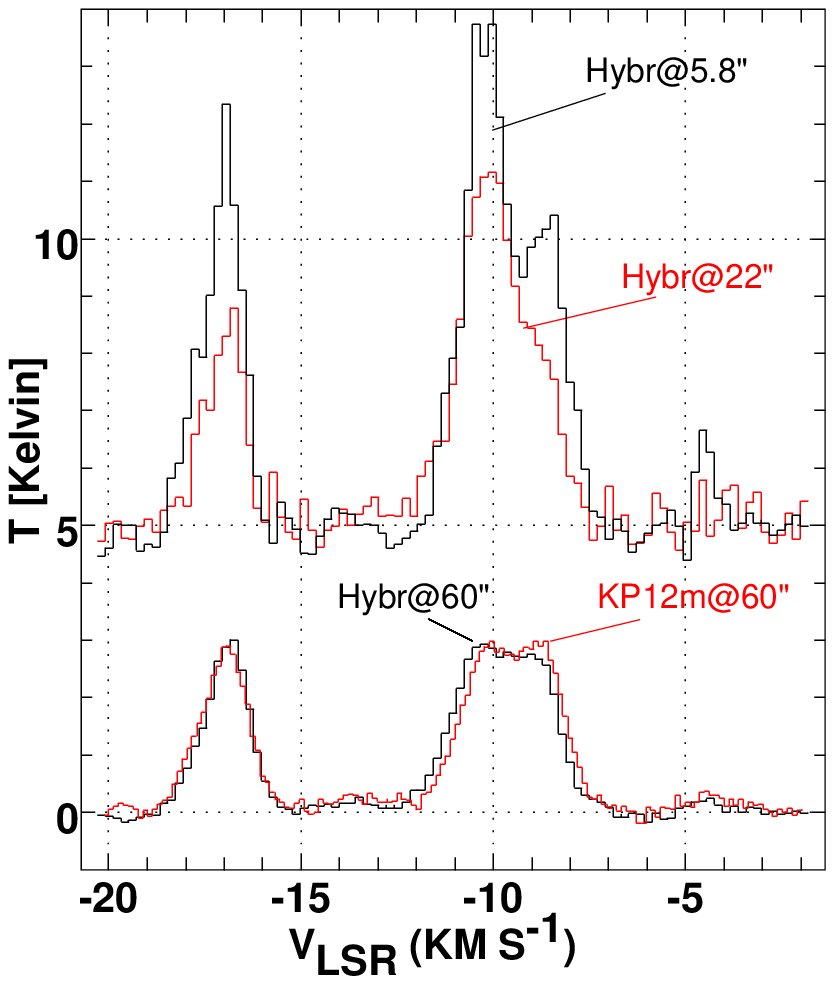}
    \caption{Evolution of J=1-0 \twCO\ brightness with angular resolution.
      At bottom, the profile from the former NRAO 12m telescope is compared
      with hybrid synthesis data convolved to 1\arcmin\ HPBW.  At top the
      hybrid data centered on the continuum source are displayed at full
      (5.8\arcsec) and 30m (22.5\arcsec) resolution.  Note that the hybrid
      synthesis profile much more closely resembles the absorption spectra
      shown in Fig.~\ref{fig:B0355f1} at bottom.}
    \label{fig:ProfileEvolution}
  \end{figure}}

\newcommand{\FigAvgMoments}{%
  \begin{figure*}
    \begin{center}
      \includegraphics[width=0.91\hsize{}]{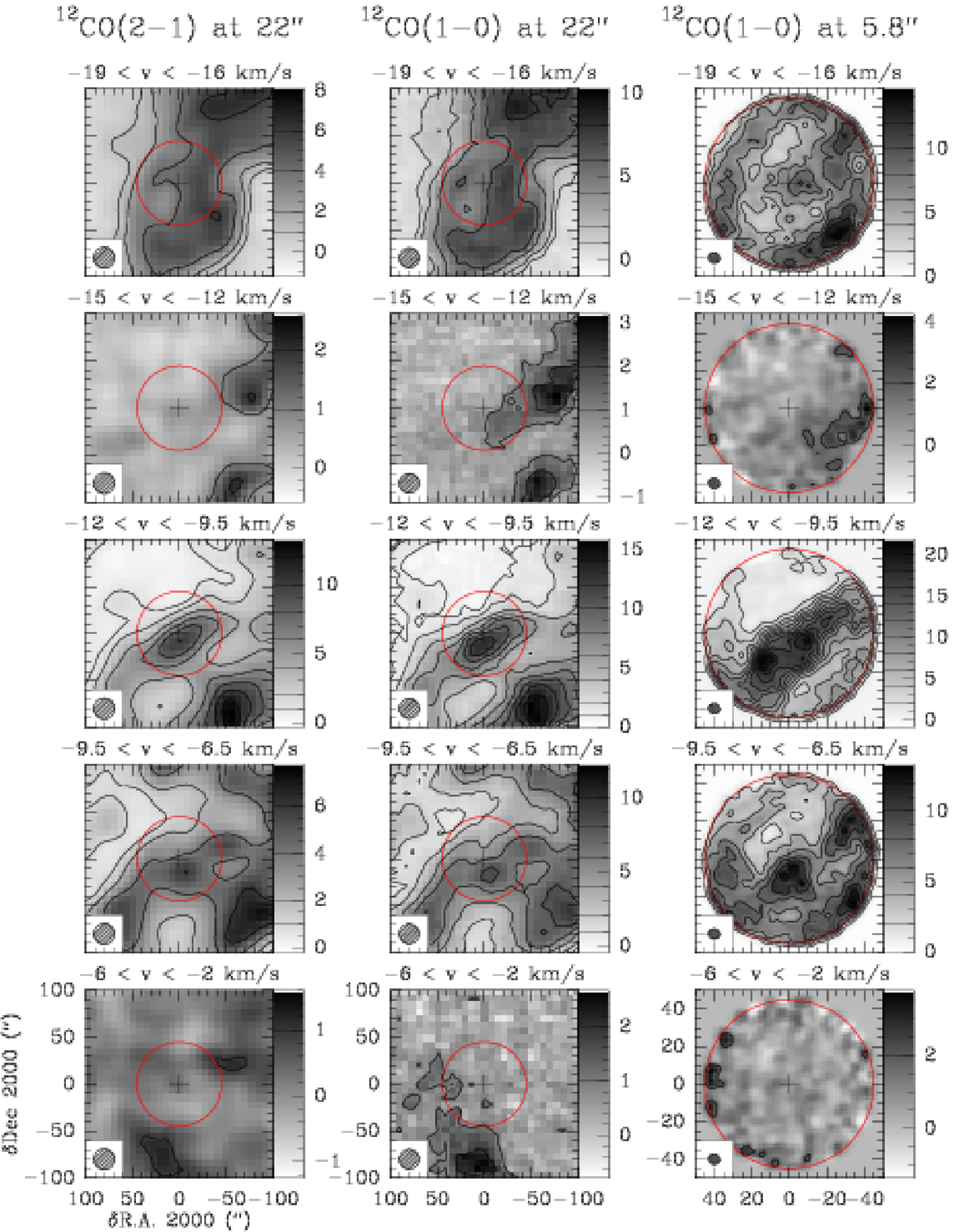}
    \end{center}
    \caption{Summary of mapping information of the diffuse gas around
      NRAO150. Each panel shows the \twCO{} emission integrated over the
      velocity range of one absorption feature visible in
      Fig.~\ref{fig:B0355f1}, \ie{} from top to bottom the $-17.5$, $-14$,
      $-10.5$, $-8.5\kms$ and, combined, the $-4.5$ and $-3.5\kms$
      features.  The position of NRAO150 is marked by a cross at the map
      center. The left and middle columns show respectively the
      \twCO{}~\Jtwo{} and \Jone{} emission observed with the 30m singledish
      antenna while the right column shows the \twCO{}~\Jone{} emission at
      higher resolution from hybrid synthesis data (\ie{} the combination
      of the 30m and PdBI instruments). The 30m \twCO{}~\Jtwo{} map has
      been smoothed to $22.5\arcsec$. The field of view of the hybrid
      synthesis maps is about half that of the singledish data: As a guide,
      the red circle has the same 45\arcsec{}-radius in all panels. The
      resolution of the hybrid synthesis data (5.8\arcsec) is almost 4
      times better than that of the singledish data, 22.5\arcsec.  Contour
      levels are spaced by 2 \Kkms{}, but for the first contour for the
      singledish maps which is at 1\Kkms{}.  Contour levels are shown on
      the intensity look-up tables.}
    \label{fig:AverageMoments}
  \end{figure*}}

\newcommand{\FigProfileBrightSpot}{%
  \begin{figure}
    \includegraphics[width=0.94\hsize{}]{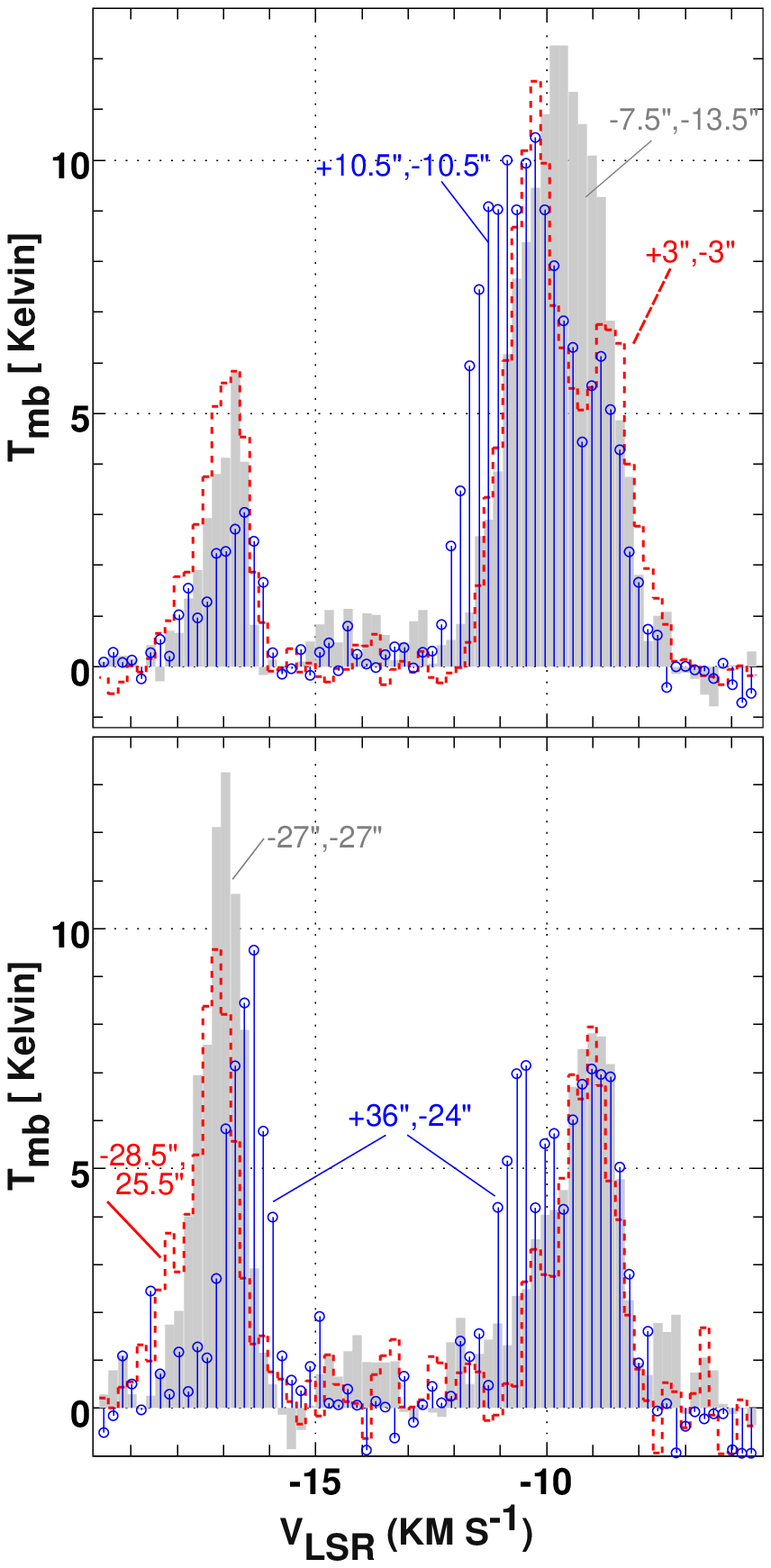}
    \caption{Hybrid synthesis CO J=1-0 (5.8\arcsec\ resolution) 
      profiles at selected locations of particularly strong emission.  {\bf
        Top:} Profiles selected from a map of peak emission near $-10\kms$.
      {\bf Bottom:} Selected from a map of peak emission near $-17\kms$.
      Profiles are labelled by their displacement $(\Delta \alpha, \Delta
      \delta)$ in arcseconds relative to the continuum source.}
    \label{fig:ProfileBrightSpot}
  \end{figure}}

\newcommand{\FigFractionalArea}{%
  \begin{figure}
    \includegraphics[height=0.94\hsize{},angle=270]{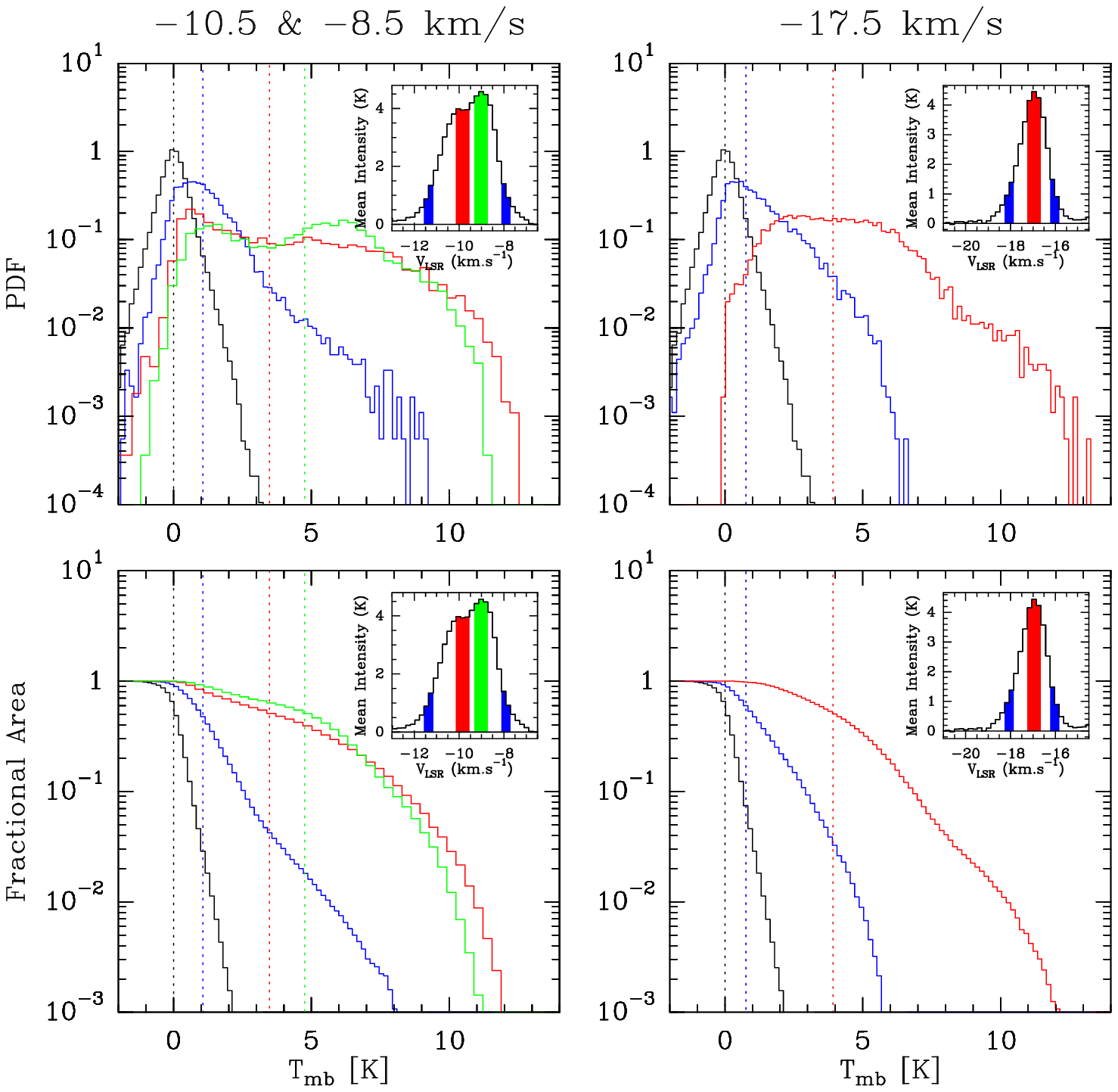}
    \caption{{\bf Top:} Probability distribution function of the line brightness 
      in the hybrid synthesis, normalized to unit area. {\bf Bottom:}
      Fractional area (cumulative distribution) of pixels whose line
      brightness is higher than the value given on the abscissa. {\bf
        Right:} for the component at $-17.5\kms$. {\bf Left:} for the
      $-10.5$ and $-8.5\kms$ components. The histograms have been computed
      in the line wings (blue curves), the line cores (red and green
      curves) and baseline devoid of signal (black curve). For reference,
      the averaged spectra are shown as inset with the same color scheme.
      The vertical dotted lines display the line brightness medians.}
    \label{fig:PixelHistograms}
  \end{figure}}

\newcommand{\FigExcessRMS}{%
  \begin{figure}
    \includegraphics[width=\hsize]{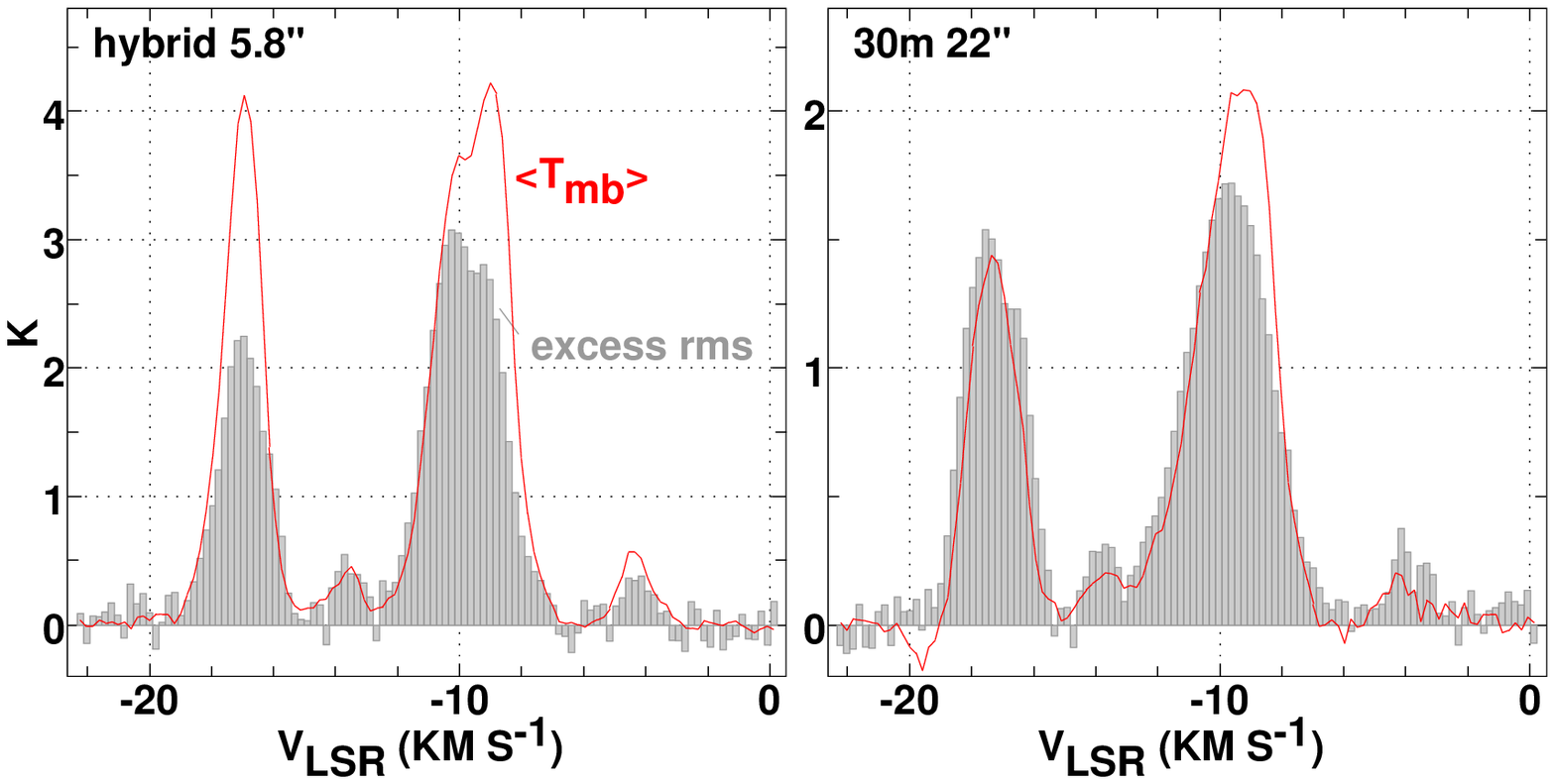}
    \caption{Mean profiles ($<\Tmb>$, red lines) and rms spatial
      brightness fluctuation in excess of the random noise rms
      ($\sqrt{\mbox{RMS}(\Tmb)^2-\mbox{RMS(noise)}^2}$, shaded histrogram).
      {\bf Left:} For the 5.8\arcsec\ resolution hybrid synthesis data
      (random noise: 0.47\K{}). {\bf Right:} For the 30m OTF mapping data
      at $22.5''$ resolution (random noise: 0.33\K{}).}
    \label{fig:ExcessRMS}
  \end{figure}}

\newcommand{\FigKine}{%
  \begin{figure*}
    \begin{minipage}{0.5\hsize}
      \begin{center}
        \includegraphics[width=0.9\hsize{}]{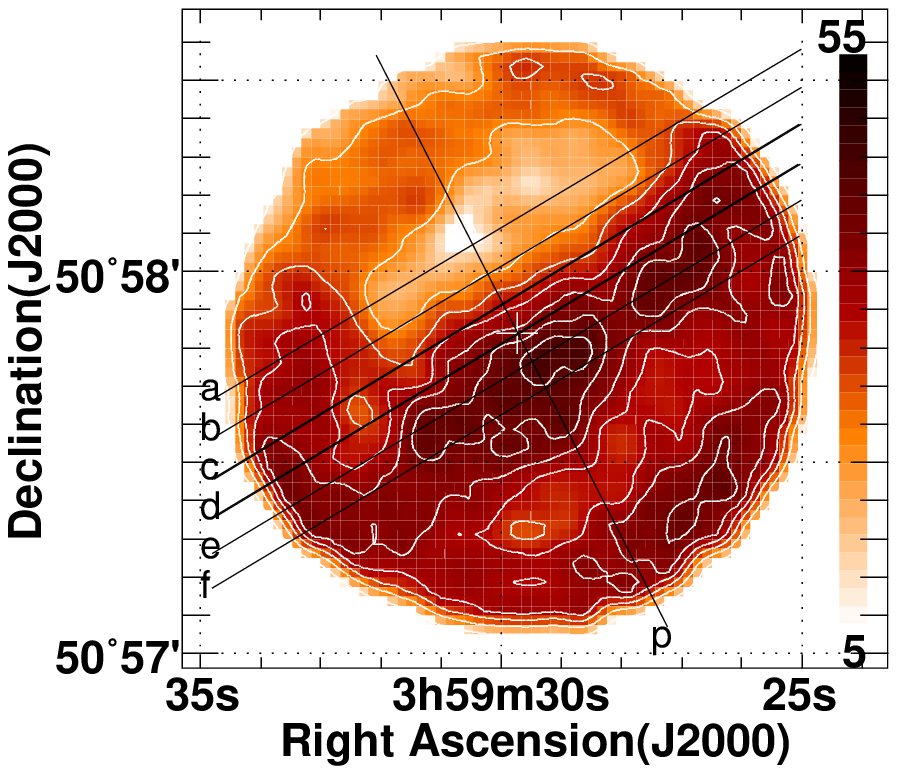}
        \includegraphics[width=0.9\hsize{}]{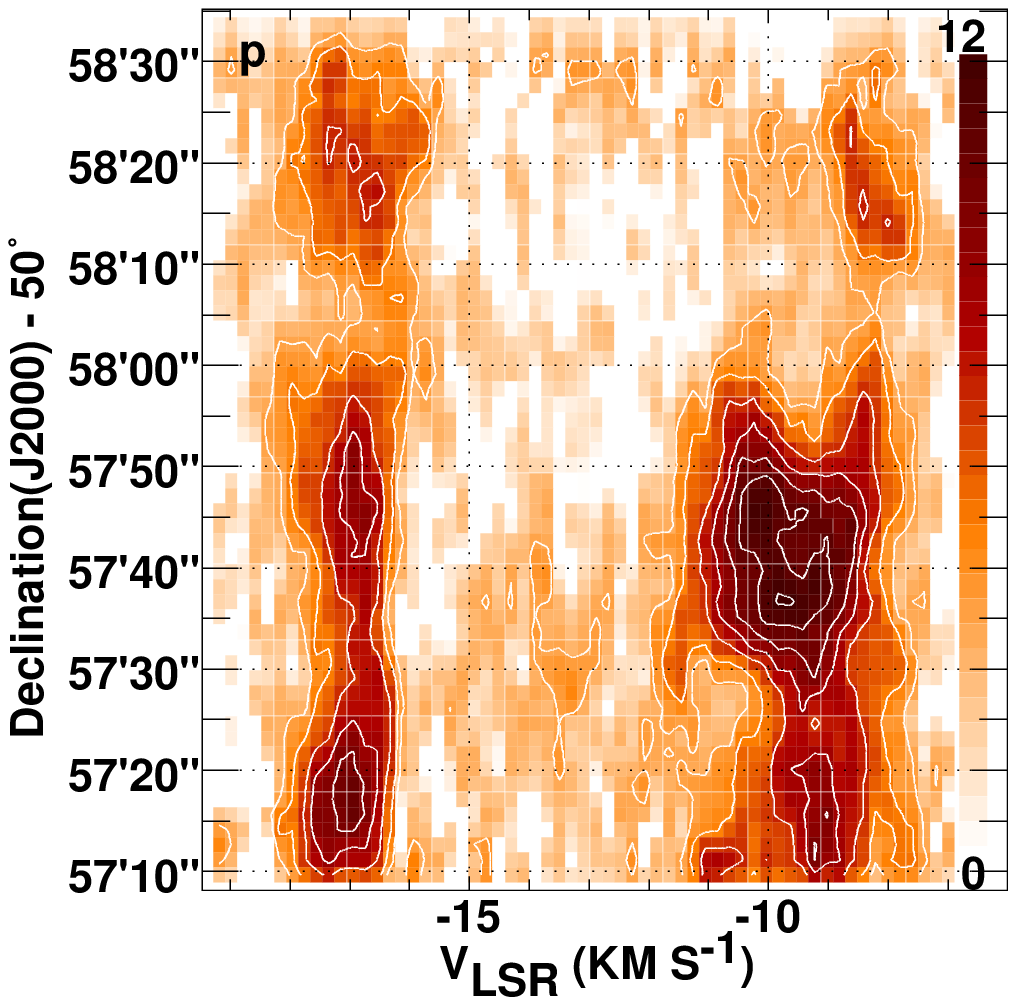}
      \end{center}
    \end{minipage}
    \begin{minipage}{0.5\hsize}
      \begin{center}
        \includegraphics[width=0.95\hsize{}]{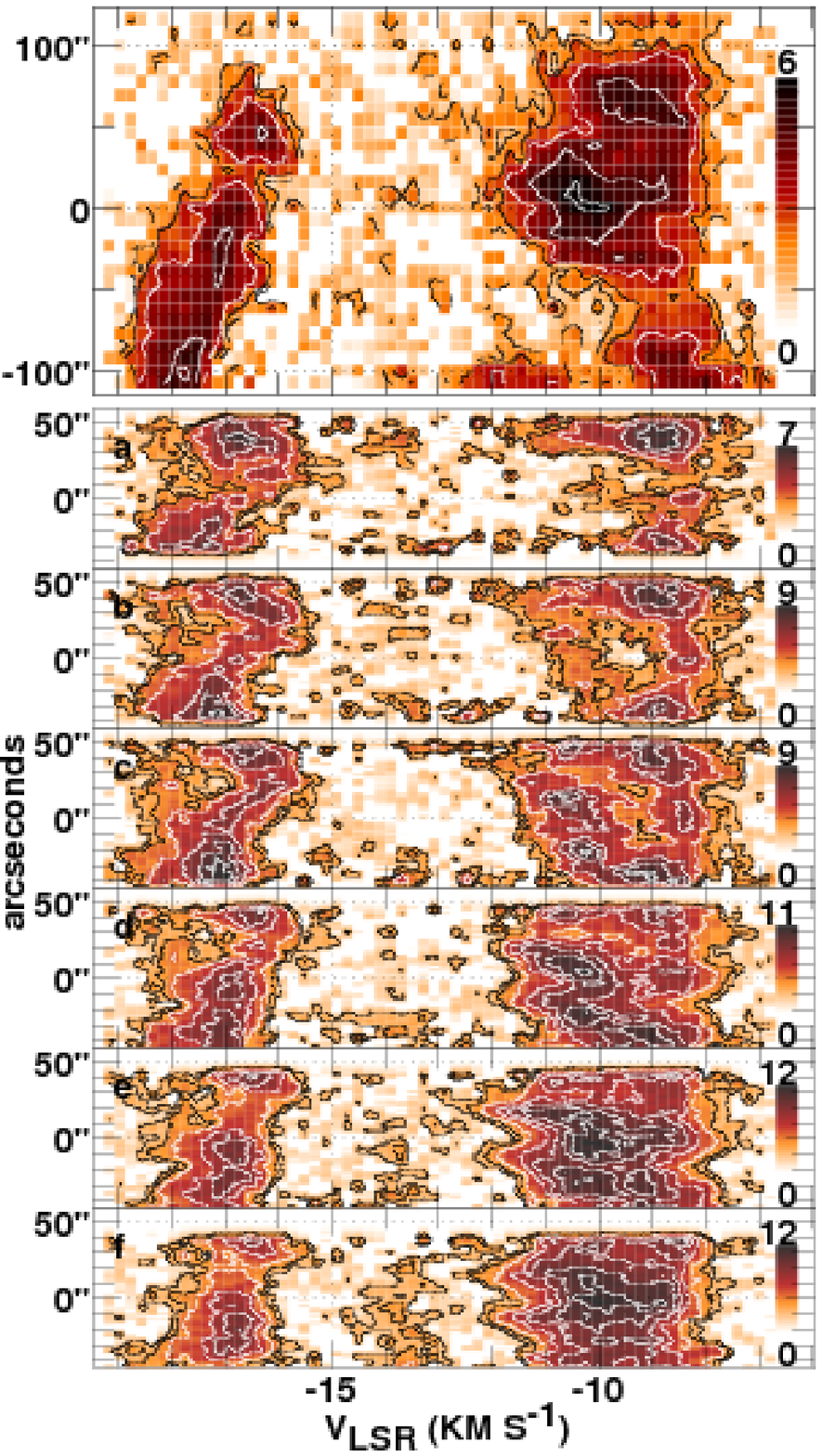}
      \end{center}
    \end{minipage}
    \caption{Position--velocity diagrams across and along the main trunk
      feature at a position angle of $-50$\degr. {\bf Top, left:}
      Integrated intensity map of 5.8\arcsec-resolution hybrid synthesis
      data summed over the disjoint velocity ranges $[-19..-15]\kms$ and
      $[-12.5..-7.5]\kms$. Contours are shown at levels 5,10,15,..35\K
      \kms{}.  The position of NRAO150 is marked by a cross at the map
      center. {\bf Bottom, left}: A position--velocity diagram along {\bf
        the perpendicular strip p through the position of NRAO150} at
      5.8\arcsec\ resolution. Contours are shown at levels
      0.5,1,2,4,6,8,10,12\K{}.  {\bf Top, right}: Longer position--velocity
      diagram at 22.5\arcsec\ resolution straddling strip d, made from 30m
      data.  Contours are shown at levels 0.5,1,2,4,6,8,10,12 \Kkms. {\bf
        Bottom, right}: Position--velocity maps at 5.8\arcsec\ resolution
      along strips a-f.  Coordinates are displacement in right ascension
      relative to NRAO150.  Strip d precisely crosses the continuum
      source.}
    \label{fig:Kine}
  \end{figure*}}

\newcommand{\FigCOcorrelation}{%
  \begin{figure*}
    \includegraphics[height=0.94\hsize{},angle=270]{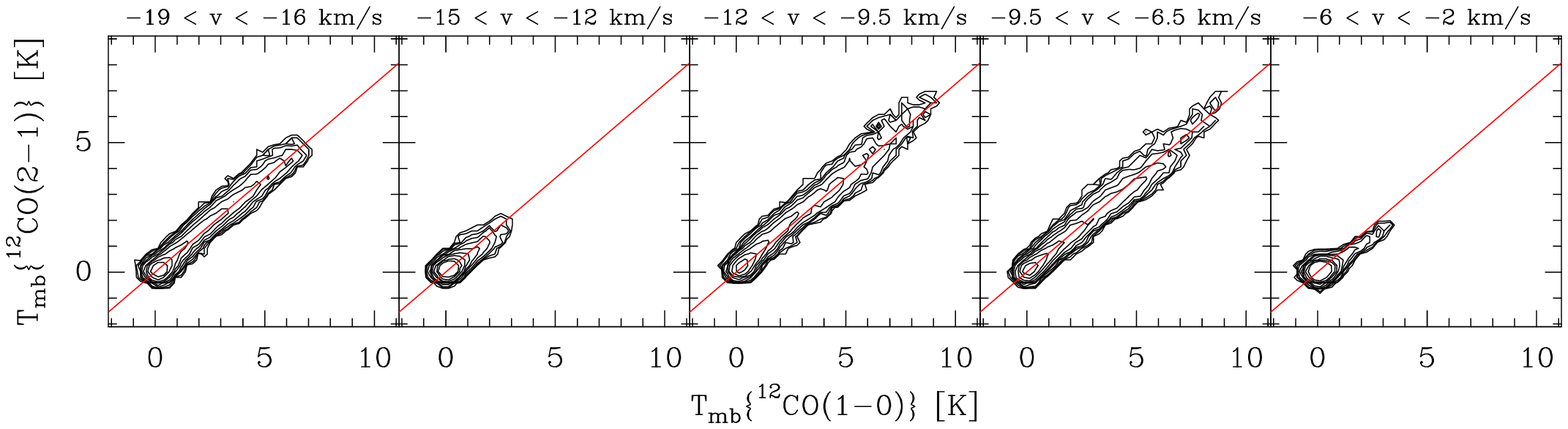}
    \caption{Joint distributions of the emission of \twCO{}~\Jtwo{}
      and~\Jone{}. The value at a given position of these joint histograms
      is the number of pixels of the input images whose intensities lye in
      the respective vertical and horizontal bins. Contour levels are set
      to 1, 2, 4, 8, 16, ... 512 points per pixel. The joint distributions
      have been computed and plotted for each velocity components. A (red)
      line of slope of 0.725 is shown for reference.  Only the emission
      around $v = -4\kms$ in the rightmost panel departs noticeably from
      the mean behavior.}
    \label{fig:12co21-vs-12co10}
  \end{figure*}}

\newcommand{\FigModelTBvsTB}{%
  \begin{figure}
    \includegraphics[width=0.94\hsize{}]{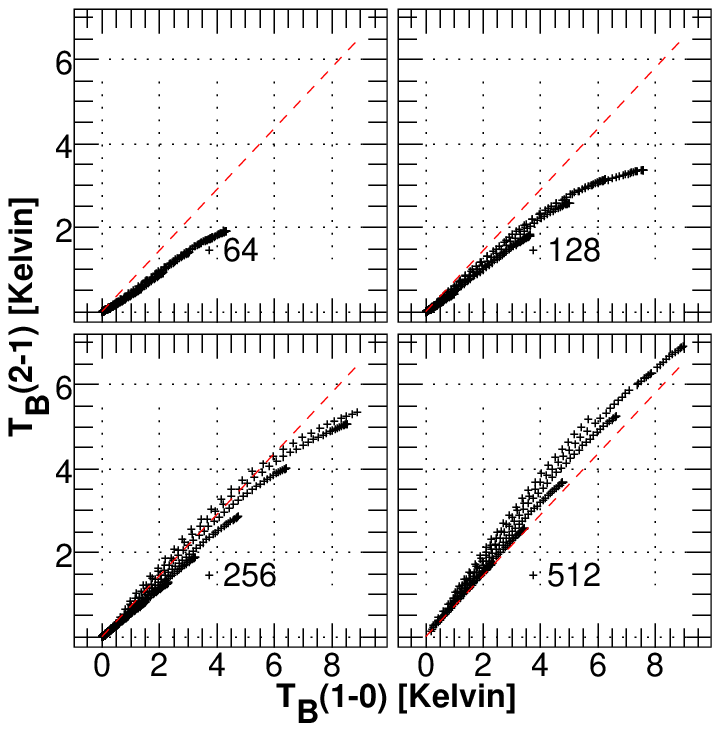}
    \caption{Non-LTE models of CO brightness.  The kinetic temperature,
      molecular hydrogen and carbon monoxide abundance are
      self-consistently determined for gas spheres of constant total
      density n(H) = n(H I) + 2 n(\HH) = 64 .. 512\pccm{} as labelled.  The
      CO line brightnesses are found from microturbulent radiative transfer
      across the faces of individual models. Each series of points shows
      the result for a different central column density varying in steps of
      2 over the range $2\times 10^{20} \pscm \le N(\mbox{H}) \le 4\times
      10^{21} \pscm$.  A slope of 0.725 is shown by a dashed (red) line.}
    \label{fig:ModelTBvsTB}
  \end{figure}}

\newcommand{\FigThmMoments}{%
  \begin{figure*}
    \includegraphics[width=0.94\hsize{}]{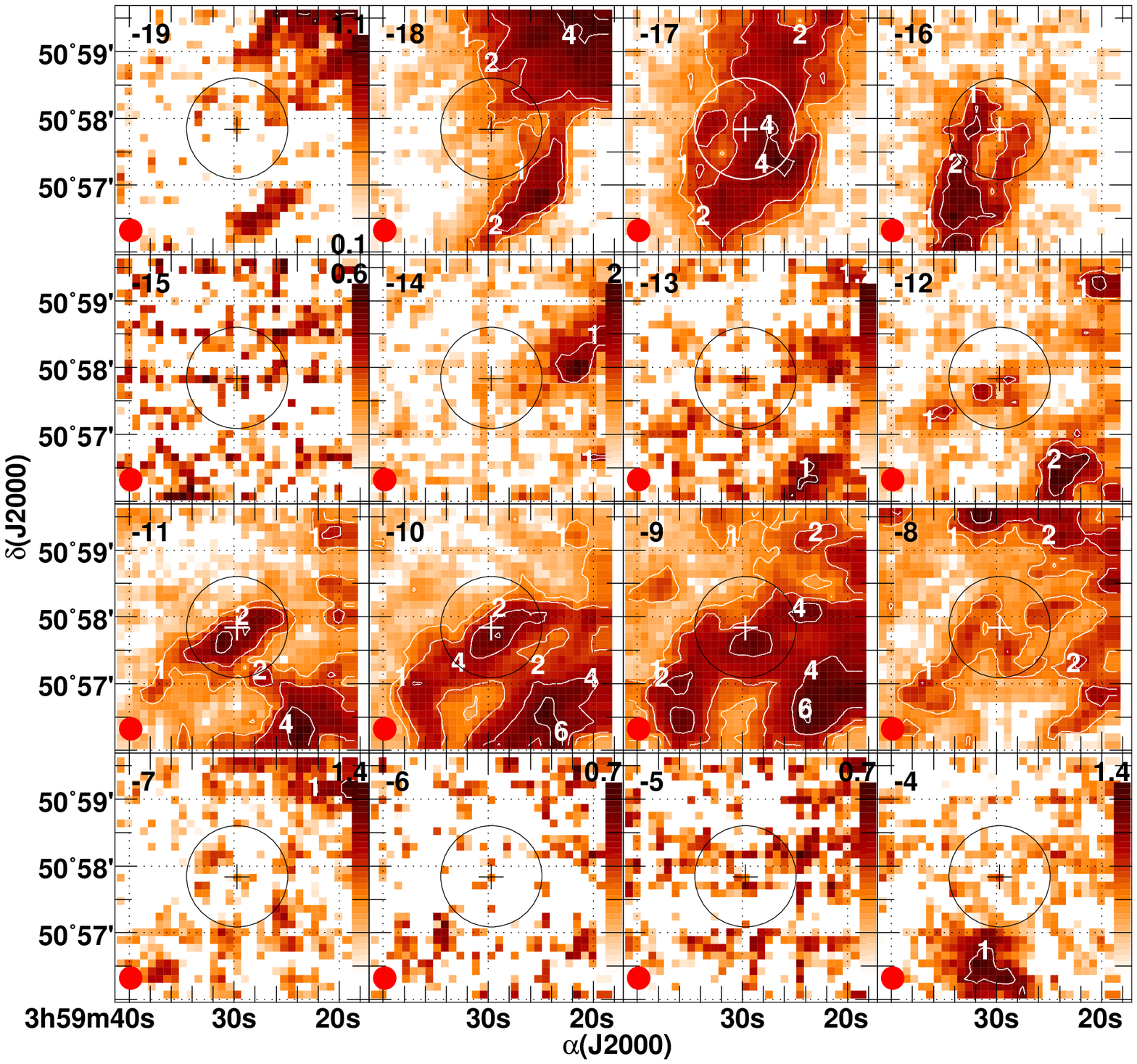}
    \caption{Finer-scale singledish maps of integrated \twCO\ brightness.  
      The 30m on-the-fly mapping observations used as short-spacing input
      to the hybrid synthesis are summed over the 5-channel (1.015 \kms)
      intervals most closely centered about the velocity indicated at the
      upper left in each panel and the integrals are scaled down by 1/1.015
      to reflect mean brightness over a 1 \kms-wide interval.  Contours are
      shown at levels 1,2,4,6,8, 10\Kkms.  The position of NRAO150 is
      marked by a cross at the map center and the region synthesized is
      shown as a circle.  Note the velocity gradient in the $-17\kms$
      feature across the position of NRAO150.  The 22.5\arcsec\ HPBW is
      shown inset at the lower left in each panel.}
    \label{fig:30mMoments}
  \end{figure*}}

\newcommand{\FigHybMoments}{%
  \begin{figure*}
    \includegraphics[width=\hsize{}]{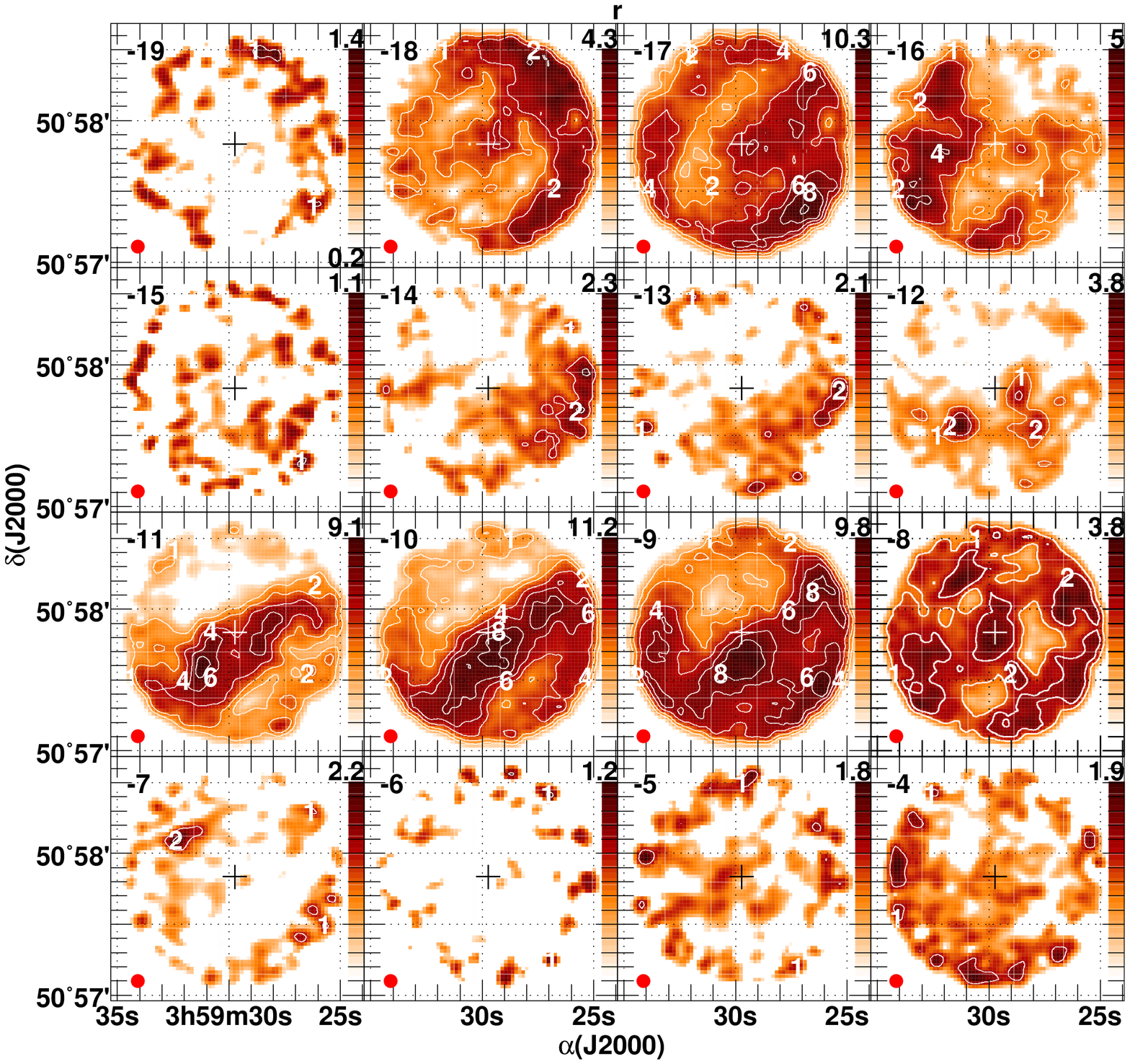}
    \caption{Finer-scale hybrid synthesis maps of integrated \twCO\ brightness.  
      The 5.8\arcsec\ hybrid synthesis profiles are summed over the
      5-channel (1.05 \kms) intervals most closely centered about the
      velocity indicated at the upper left in each panel and the integrals
      are scaled by 1/1.05 to reflect mean brightness over a 1\kms-wide
      interval.  Contours are shown at levels 1,2,4,6 and 8\Kkms. The
      position of NRAO150 is marked by a cross at the map center.  The
      5.8\arcsec\ synthesized HPBW is shown inset at the lower left in each
      panel.}
    \label{fig:HybrMoments}
  \end{figure*}}

\newcommand{\FigFractionalAreaTwo}{%
  \begin{figure*}
    \centering %
    \includegraphics[height=0.36\hsize{},angle=270]{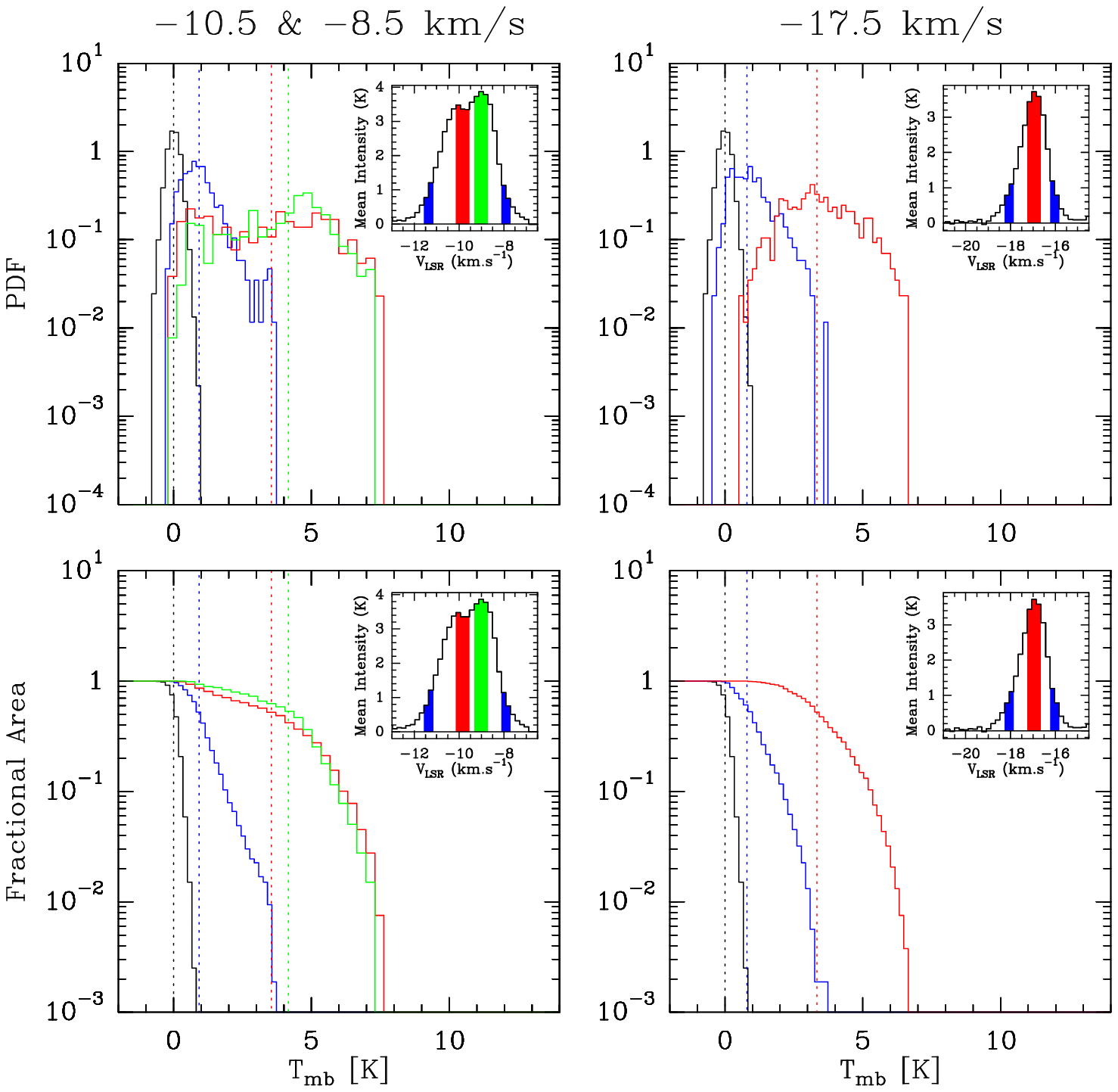}
    \hspace{0.5cm}
    \includegraphics[height=0.36\hsize{},angle=270]{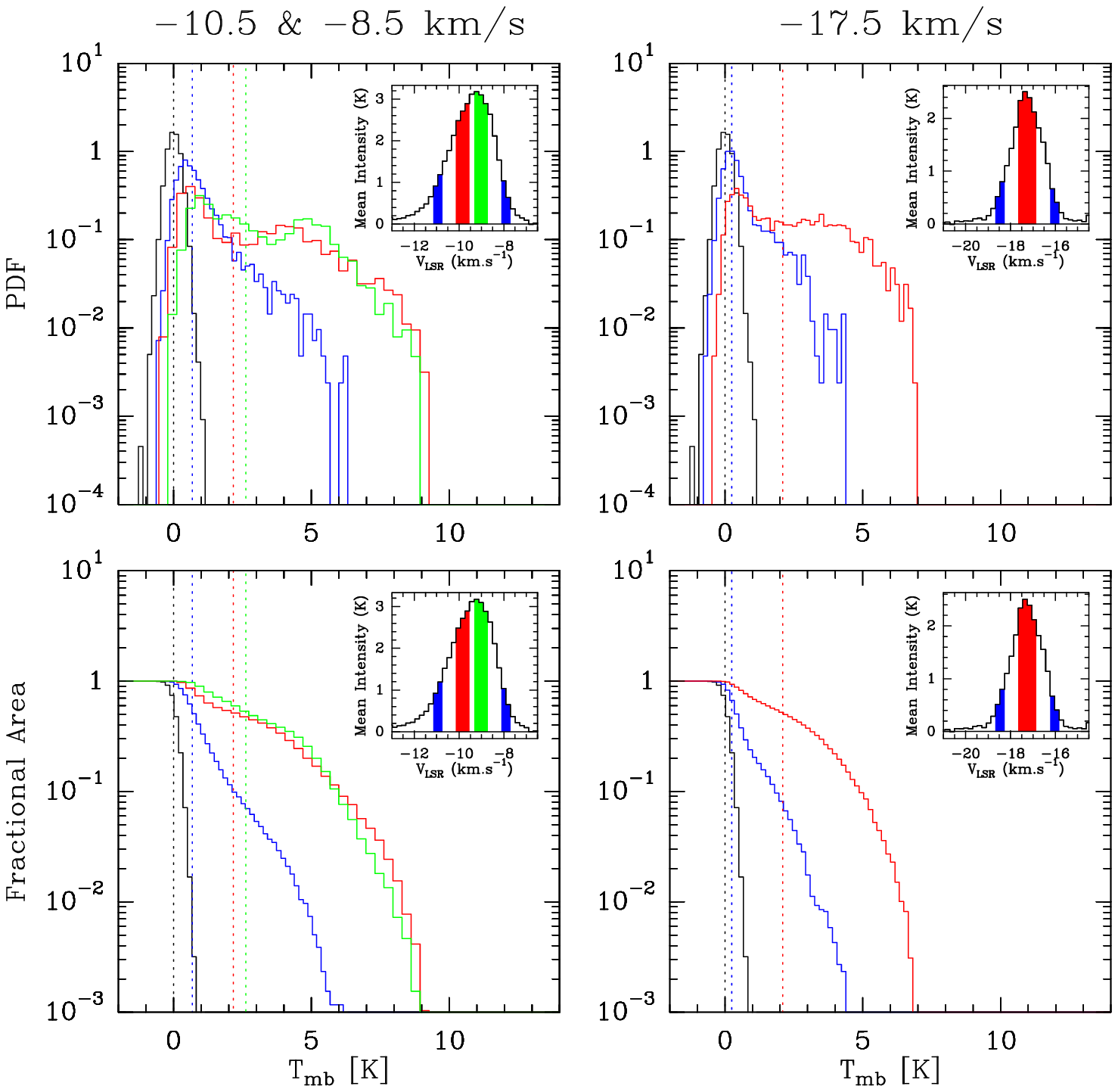}
    \caption{{\bf Left:} Same as Fig.~\ref{fig:PixelHistograms} but using 
      the 30m data only. {\bf Right:} Same as right but using the total
      field-of-view of the 30m data, \ie{} $200''\times 200''$.}
    \label{fig:FracArea:2}
  \end{figure*}}

\newcommand{\FigMoments}{%
  \begin{figure*}
    \centering %
    \includegraphics[height=0.91\hsize{},angle=270]{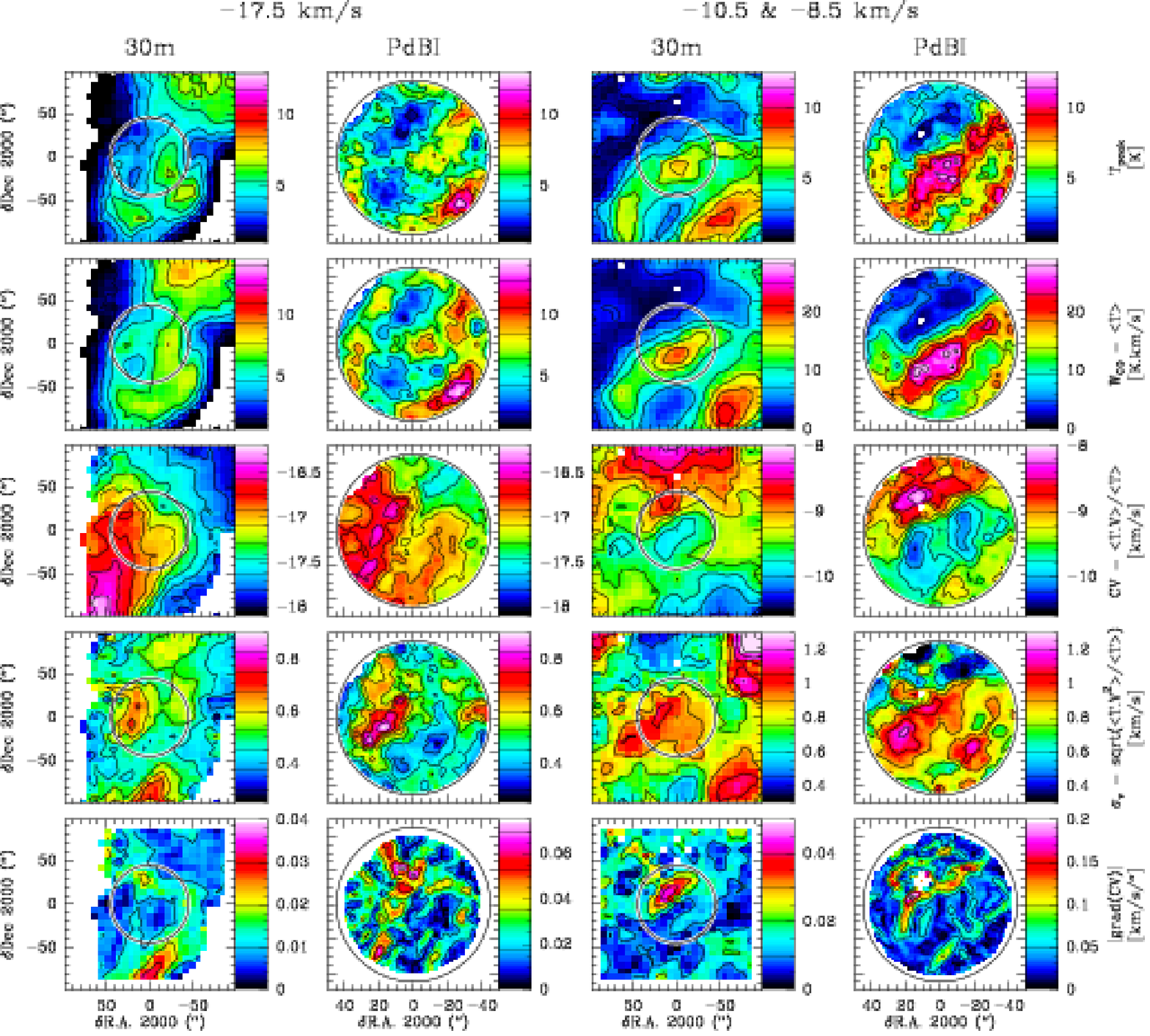}
    \caption{Moments of the components $-17.5\kms$ and $-10.5$ and $-8.5\kms$
      together observed at two different resolution. From top to bottom:
      Peak temperature, Integrated area, Centroid velocity, 2nd order
      moment (proportional to the FWHM for a Gaussian line) and the modulus
      of the gradient of centroid velocity (for a flat rotation curve of
      the Galaxy, $-20\kms$ corresponds to 2.5\kpc{} in the direction of
      NRAO150 and at 2.5\kpc{}, $0.1\kms/''$ corresponds to $8\kms \,
      \mbox{pc}^{-1}$).}
    \label{fig:Moments}
  \end{figure*}}

\newcommand{\FigBrightSpot}{%
  \begin{figure*}
    \begin{minipage}{0.5\hsize}
      \begin{flushright}
        \includegraphics[width=0.85\hsize{},angle=270]{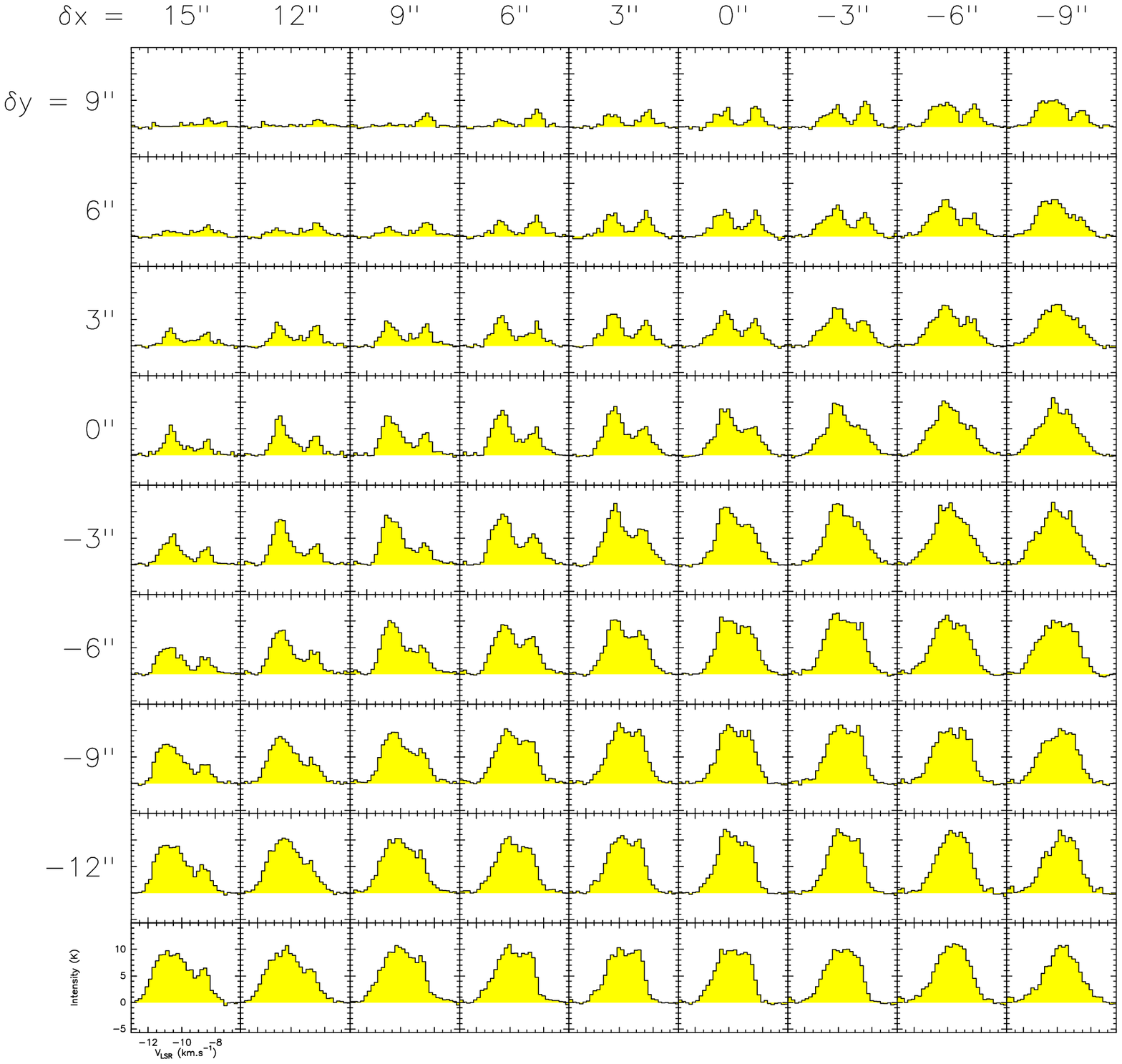}\\[\medskipamount]
        \includegraphics[width=0.85\hsize{},angle=270]{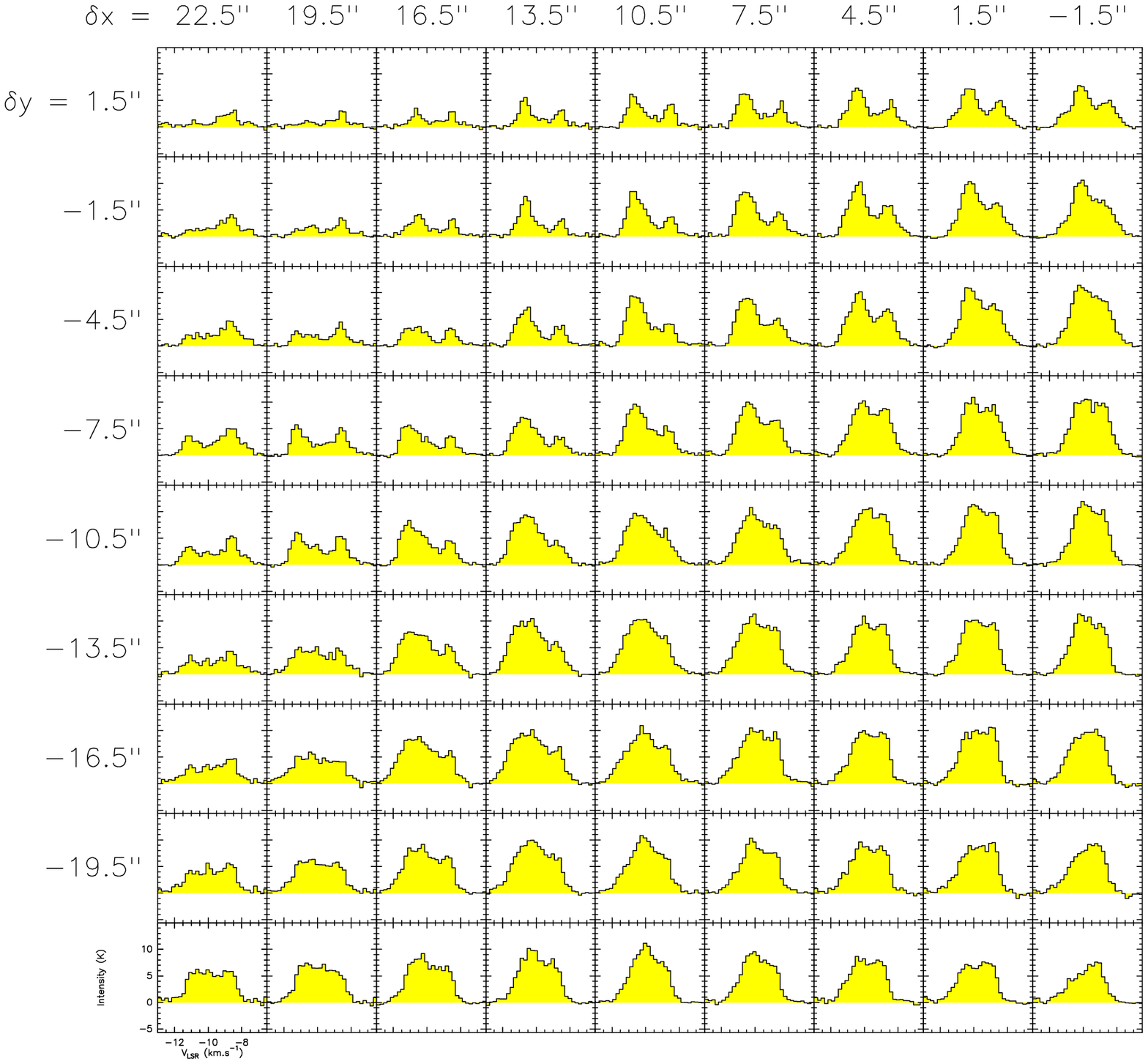}\\[\medskipamount]
        \includegraphics[width=0.85\hsize{},angle=270]{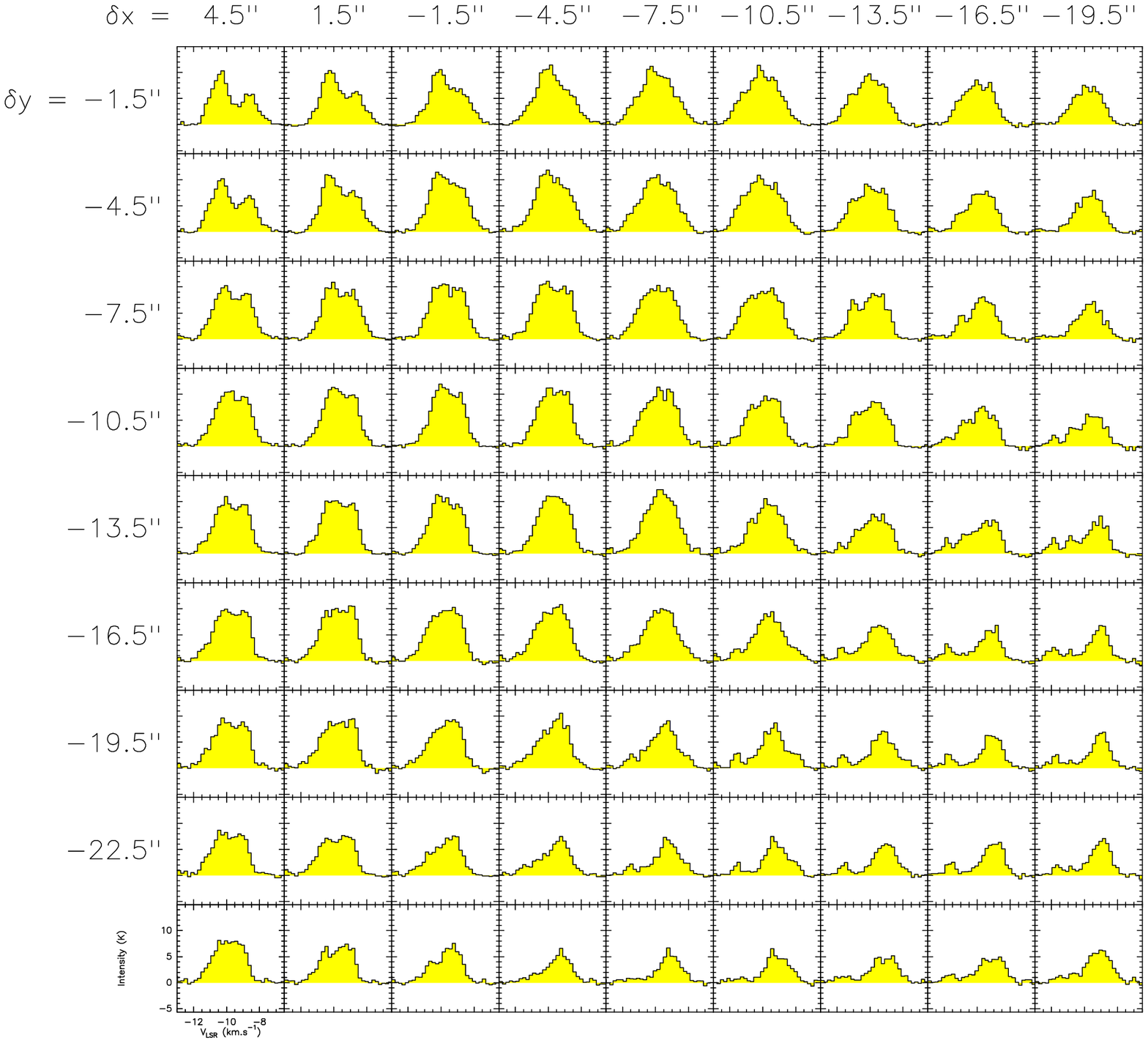}
      \end{flushright}
    \end{minipage}
    \begin{minipage}{0.5\hsize}
      \begin{flushright}
        \includegraphics[width=0.85\hsize{},angle=270]{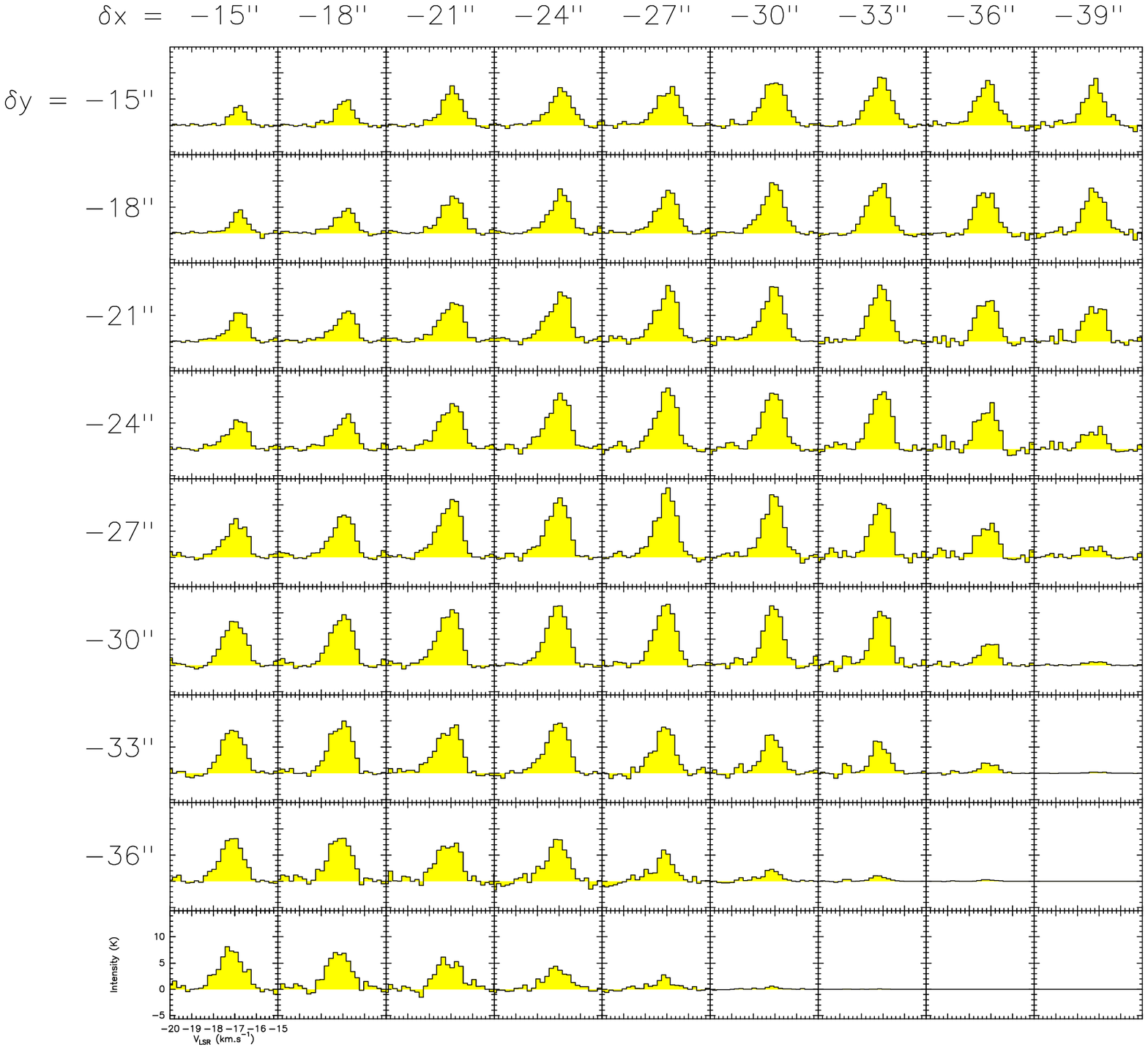}\\[\medskipamount]
        \includegraphics[width=0.85\hsize{},angle=270]{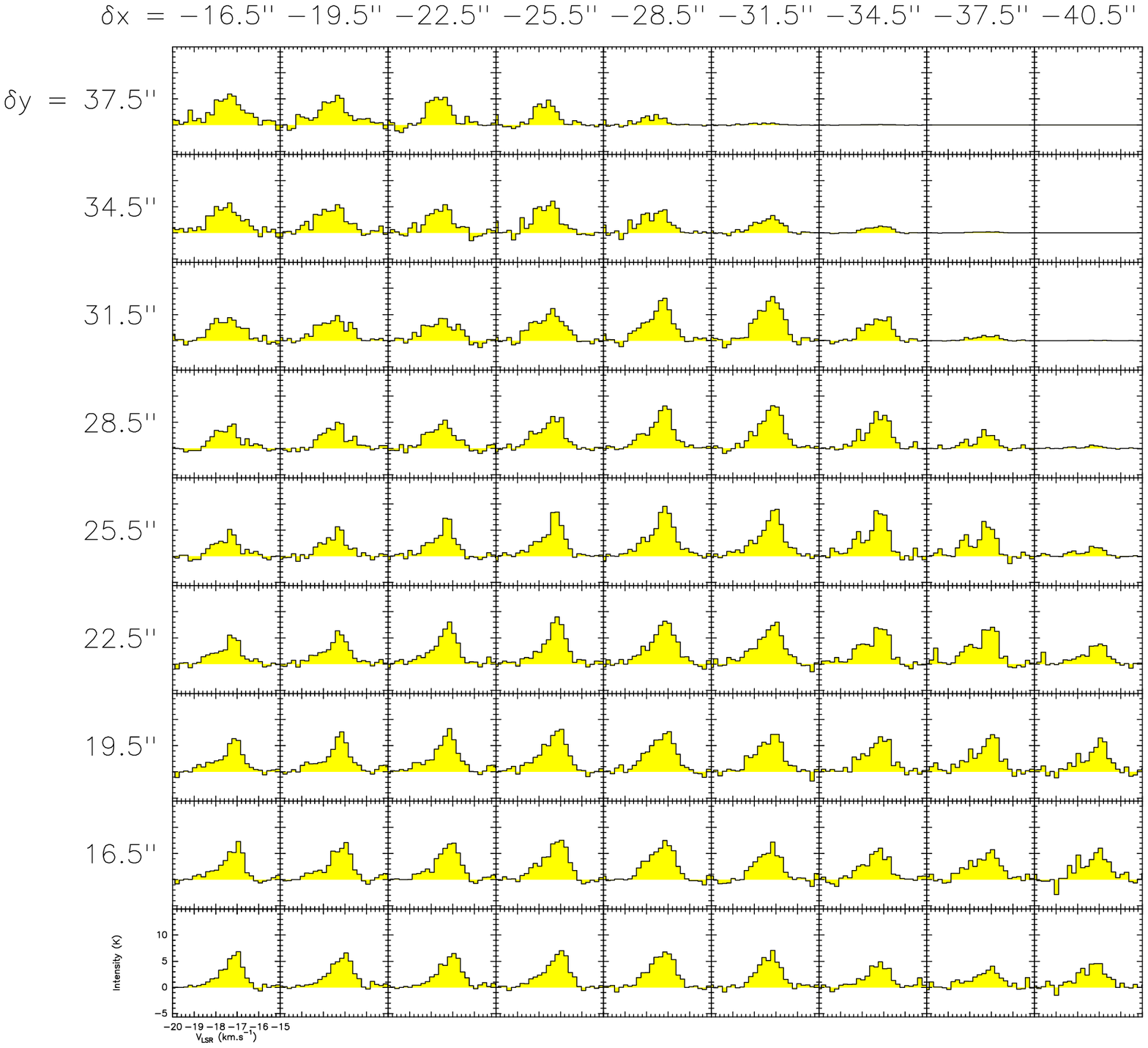}\\[\medskipamount]
        \includegraphics[width=0.85\hsize{},angle=270]{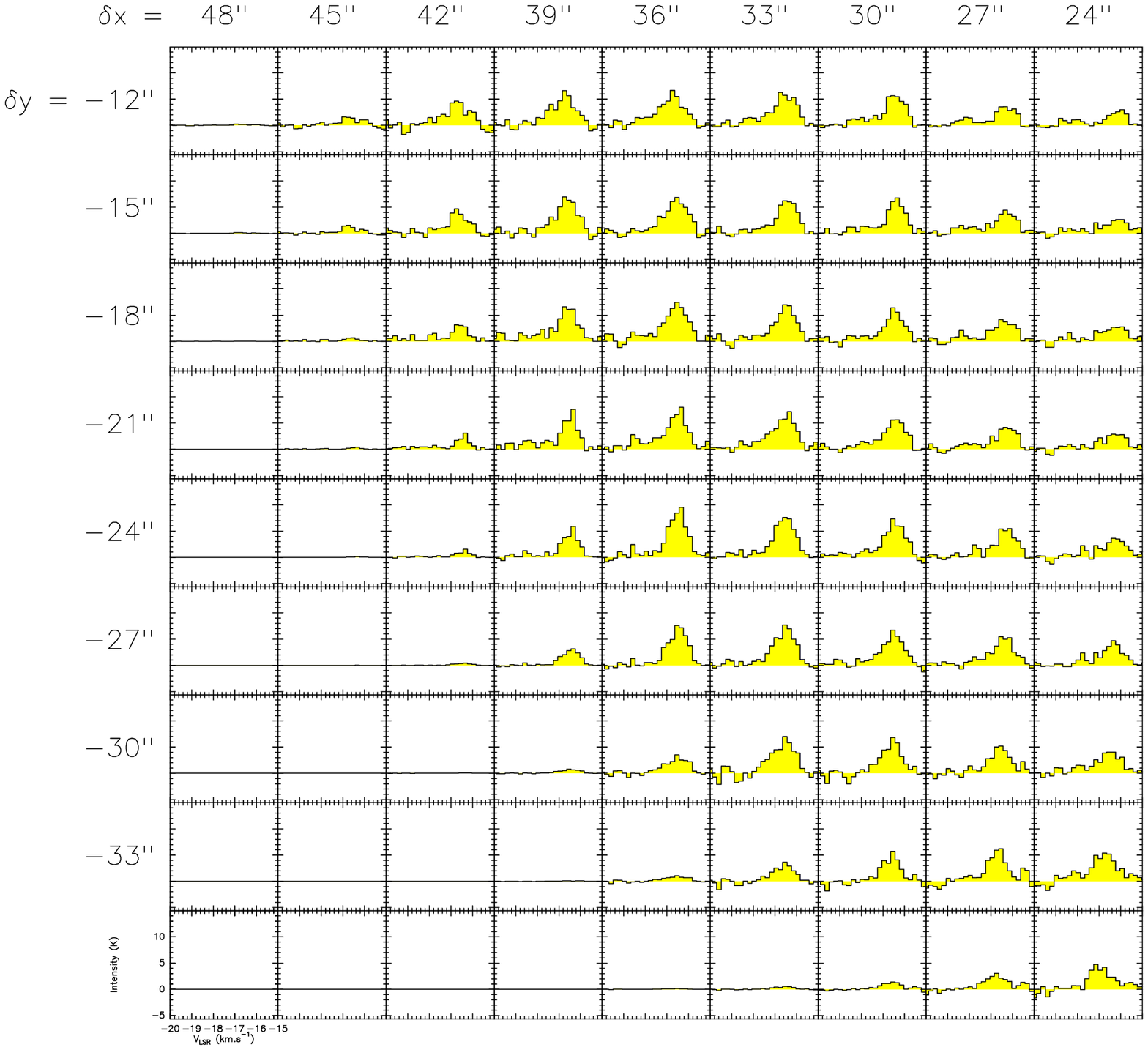}
      \end{flushright}
    \end{minipage}
    \caption{Maps of spectra around the profiles shown in 
      Fig.~\ref{fig:ProfileBrightSpot} sorted by distance to NRAO150 from
      top to bottom. {\bf Left:} The $-17.5\kms$ feature shown around
      $(+3'',-3'')$, $(+10.5'',-10.5'')$ and $(-7.5'',-13.5'')$. {\bf
        Right:} The $-10.5$ and $-8.5\kms$ features shown around
      $(-27'',-27'')$, $-28.5'',+25.5''$ and $(+36'',-24'')$.}
    \label{fig:BrightSpot}
  \end{figure*}}

\begin{document}

\title{Imaging galactic diffuse gas: \\Bright, turbulent CO surrounding the
  line of sight to NRAO150 \thanks{Based on observations obtained with the
    IRAM Plateau de Bure interferometer and 30~m telescope. IRAM is
    supported by INSU/CNRS (France), MPG (Germany), and IGN (Spain).}}

\titlerunning{Bright, turbulent carbon monoxide} %

\author{J. Pety\inst{1,2}, R. Lucas\inst{1}, and H. S. Liszt\inst{3}} %

\institute{%
  Institut de Radioastronomie Millim\'etrique, 300 Rue de la Piscine,
  F-38406 Saint Martin d'H\`eres, France \email{pety@iram.fr,lucas@iram.fr}%
  \and Obs. de Paris, 61 av. de l'Observatoire, 75014, Paris, France %
  \and National Radio Astronomy Observatory, 520 Edgemont Road,
  Charlottesville, VA, USA 22903-2475 \email{hliszt@nrao.edu}}

\date{received \today} %

\offprints{pety@iram.fr}

\abstract
{}
{To understand the environment and extended structure of the host galactic
  gas whose molecular absorption line chemistry, we previously observed
  along the microscopic line of sight to the blazar/radiocontinuum source
  NRAO150 ($aka$ B0355+508).}
{We used the IRAM 30m Telescope and Plateau de Bure Interferometer to make
  two series of images of the host gas: i) 22.5\arcsec\ resolution
  single-dish maps of \twCO\ J=1-0 and 2-1 emission over a 220\arcsec by
  220\arcsec\ field; ii) a hybrid (interferometer+singledish) aperture
  synthesis mosaic of \twCO\ J=1-0 emission at 5.8\arcsec\ resolution over
  a 90\arcsec-diameter region.\thanks{The three spectra cubes (RA, Dec,
    Velocity) are available in electronic form (FITS format) at the CDS.}}
{At 22.5\arcsec\ resolution, the CO J=1-0 emission toward NRAO150 is
  30-100\% brighter at some velocities than seen previously with 1\arcmin\ 
  resolution, and there are some modest systematic velocity gradients over
  the 220\arcsec\ field. Of the five CO components seen in the absorption
  spectra, the weakest ones are absent in emission toward NRAO150 but
  appear more strongly at the edges of the region mapped in emission. The
  overall spatial variations in the strongly emitting gas have Poisson
  statistics with rms fluctuations about equal to the mean emission level
  in the line wings and much of the line cores. The J=2-1/J=1-0 line ratios
  calculated pixel-by-pixel cluster around 0.7.  At 6\arcsec\ resolution,
  disparity between the absorption and emission profiles of the stronger
  components has been largely ameliorated. The \twCO\ J=1-0 emission
  exhibits i) remarkably bright peaks, $\Tmb = 12-13\K$, even as $4''$ from
  NRAO150; ii) smaller relative levels of spatial fluctuation in the line
  cores, but a very broad range of possible intensities at every velocity;
  and iii) striking kinematics whereby the monotonic velocity shifts and
  supersonically broadened lines in 22.5\arcsec\ spectra are decomposed
  into much stronger velocity gradients and abrupt velocity reversals of
  intense but narrow, probably subsonic, line cores.}
{CO components that are observed in absorption at a moderate optical depth
  (0.5) and are undetected in emission at $1'$ resolution toward NRAO 150
  remain undetected at $6''$ resolution. This implies that they are not a
  previously-hidden large-scale molecular component revealed in absorption,
  but they do highlight the robustness of the chemistry into regions where
  the density and column density are too low to produce much rotational
  excitation, even in CO. Bright CO lines around NRAO150 most probably
  reflect the variation of a chemical process, \ie{} the C\p-CO conversion.
  However, the ultimate cause of the variations of this chemical process in
  such a limited field of view remains uncertain.}

\maketitle{}

\section{Introduction}

Many detailed and sensitive studies of interstellar gas employ absorption
spectra taken against the microscopic disks of stars or compact radio
continuum sources.  We have recently used such spectra at radiofrequencies
to elucidate an unexpectedly widespread molecular chemistry in diffuse
clouds; for instance, see \cite{LisLuc+06} and the many references
therein).

Lamentably, observing gas in absorption along such single, isolated lines
of sight does not generally allow the identification of the intervening
features with recognizable physical bodies in space, and their possible
association with other nearby entities (for instance dark clouds) remains
hidden.  These limitations have obvious consequences for our understanding
of the intervening medium and the absorption profiles themselves.  But in
some rare cases it is becoming possible to observe the same feature in
emission and absorption, even in the same chemical species, and so perhaps
to characterize the nature of the intervening material by imaging it across
the line of sight away from the background object.

This paper begins a short series of reports of studies undertaken to
elucidate the nature of the gas occulting the compact extragalactic mm-wave
background sources employed in our work.  It describes one case, using the
J=1-0 and 2-1 mm-wave rotational transitions of carbon monoxide to image
the particularly complex kinematics of relatively distant absorbing
material along the low-latitude line of sight toward NRAO150
\citep{CoxGue+88} at the highest usefully attainable resolution.  It is
organized as follows.  Section 2 describes the sightline toward the blazar
NRAO150 as it is understood from radiofrequency observations of CO and
\hcop\ in emission and in approximately a dozen other atomic and molecular
species in absorption.  Section 3 describes the techniques used to image
and/or synthesize the pattern of foreground CO emission using singledish
and interferometeric synthesis data.  Section 4 describes the results of
the new observations and Section 5 is dedicated to models and Section 6 is
a summary.

\section{The line of sight toward the blazar NRAO150}

\FigIntro{} %

Unlike the vast majority of strong radio continuum sources, no visual
identification of the blazar NRAO150 \citep{P-TWad+66} has yet been made.
At $\alpha=3$h59m29.73s, $\delta=50$\degr 57\arcmin 50.2\arcsec\ (J2000),
NRAO150 is seen at low latitude in the outer Galaxy ($l,b=$150.3772\degr,
-1.6037\degr), so the path through the galactic disk is long and there is a
heavy accumulation of distributed diffuse/H I gas. Quite aside from any
dense material which might intervene, the integrated H I emission of the
nearest spectrum in the Leiden-Dwingeloo H I survey \citep{HarBur97} gives
an optically-thin lower limit N(H I) $ >7.5\times10^{21}~\pcc$ or $\Av >
4\mag$.  There is an approximate 5 kpc path length from the Sun out to a
galactocentric radius of 13 kpc where the neutral gas disk might end
(assuming R$_{\sun}$ = 8.5\kpc{}).  For a typical mean gas density of
$1\pccc$, this corresponds to N(H) $= 10^{22}~\pcc$ or $\Av \approx 6\mag$.

The sightline also harbors denser \HH-bearing clouds as shown in
Fig.~\ref{fig:B0355f1}. While the H I absorption spectrum is too heavily
saturated to be revealing \citep{LisLuc96}, the CO and \hcop{} absorption
spectra reveal 5-6 absorption features at LSR velocities of $-17.5$, $-14$,
$-10.5$, $-8.5$, $-4.5/-3.5\kms$. All those absorption features have
comparable N(\hcop) $\approx 1 \times 10^{12}\pscm$ implying N(H) $\approx$
2N(\HH) $\approx 2 \times 10^{21}\pscm$ \citep{LucLis96,LisLuc96} or $\Av
\approx 1\mag$.  Each cloud also has N(OH) comparable to that seen toward
\zoph{} (\ibid{}), the prototypical diffuse cloud sightline studied
optically \citep{VanBla86} where the visual extinction is 1\mag{}.  As in
optically-studied diffuse clouds \citep{SonWel+07,BurFra+07,SheRog+07}, we
derived N(CO) $< 10^{15} - 10^{16}\pscm$ \citep{LucLis98}. This implies
that only a modest fraction of the expected free gas-phase carbon N(C)
$\approx 1.6\times10^{-4}$ N(H) $\approx 3 \times 10^{17}\pscm$
\citep{SofLau+04} is incorporated in CO, as in any cloud with \AV $\la 1$
mag.

We conclude that the molecular features seen toward NRAO150 are hosted in
diffuse or marginally translucent gas, consistent with the weakness of
emission spectra in species other than CO and H I.  OH emission seen at
18\arcmin\ resolution is quite weak (0.01\K{}) and very broad, extending
over $-35<v<0\kms$ and individual cloud features cannot be discerned
\citep{LisLuc96}.  The only species seen in mm-wave emission beside CO is
\hcop\ which appeared very weakly (0.01-0.02\K{}) at $v = -9\kms$ when
averaged over a several arcminute region \citep{LucLis96}.

The absorption features display a rich and varied chemistry, a small sample
of which is illustrated in the spectra of the bottom panel of
Fig.~\ref{fig:B0355f1}. They harbor a prominent complement of typically
ubiquitous species such as CO, \cch\ and \c3h2\ \citep{LucLis00C2H} as well
as OH.  All of the clouds also have detectable absorption in HNC, although
weakly at $-14\kms$ and $-4.5/-3.5\kms$. The yet-stronger HCN spectrum is
heavily confused by hyperfine structure, see \cite{LisLuc01} for a survey
discussion of the CN-bearing species.  The features at $-17.5\kms$ and
$-10.5\kms$ are more chemically complex, with much more readily detectable
amounts of the CN-family and such sulfur-bearing species as CS, SO, \HH{}S
and HCS\p\ \citep{LisLuc02} or \hhco\ and \ammon\ 
\citep{Nas90,MarMoo+93,LisLuc95a,LisLuc+06}.

The top panel of Fig.~\ref{fig:B0355f1} shows 1\arcmin\ HPBW CO and \thCO{}
J=1-0 emission spectra from the former NRAO KP 12m telescope, which we used
(in conjunction with the absorption spectrum) to check the CO column
density and excitation temperatures toward this and other sources
\citep{LisLuc98}.  The two rather chemically-different absorption line
components at $-10.5$ and $-8.5\kms$ are heavily blended at 1\arcmin\ 
resolution, unlike the clearly double structure of the CO and other
absorption. This is atypical of our comparisons in many other lines of
sight between the absorption spectra of different species and the CO
emission profiles when the latter are so strong.  It is also curious
because, being chemically so different, it would seem reasonable to believe
that the two kinematic components were unrelated physically along the line
of sight. Alternatively, the $-10.5$ and $-8.5\kms$ components could be
related and mixed with or enveloped by warmer optically thin gas which
fills in the emission spectrum at $-10\kms$ but not the absorption.
However, as will be shown here, the two features are indeed part of the
same structure and disparity between the emission and absorption profiles
arises solely as the result of beam-smearing at 1\arcmin\ resolution.

\TabObs{} %

Several of the features (at $-14$ and $-4.5/-3.5\kms$) seen so strongly in
CO and \hcop\ absorption are almost absent in CO emission. Can these
particular clouds really be dense enough to form \hcop\ and the
hydrocarbons \cch\ and \c3h2, along with relatively small amounts of HNC
and HCN, but not dense enough to form detectable amounts of \ammon\ or
\hhco\ \citep{LisLuc+06} or even dense enough to excite CO to detectable
levels?  The clouds at $-14$, $-4.5/-3.5\kms$ which are so weak in CO
emission have non-negligible J=1-0 optical depths 0.3 - 0.6 in \twCO,
implying that their excitation temperatures are less than 1\K{} above the
cosmic microwave background.  Comparably low excitation temperatures have
recently been found to be common in CO seen in the $uv$ absorption spectra
of bright stars when N(CO) $< 2 \times 10^{15}~\pcc$ and N(\HH) $< 5 \times
10^{20}~\pcc$ \citep{BurFra+07,SonWel+07}, the latter equivalent to
N(\hcop) $\la 10^{12}~\pcc$.  Although one might also wonder if such
\HH-bearing, CO-silent clouds are just very cold, perhaps part of the dark
molecular baryonic matter which has been hypothesized to exist in the disks
of galaxies \citep{PheCom+94}, the \HH\ seen in $uv$ absorption along
comparable sightlines hosting weakly-excited CO indicates kinetic
temperatures of 40-80\K{} and the kinetic temperature inferred from C$_2$
is only slightly lower (\ibid{}).

In summary, the line of sight towards NRAO150 is very heavily extincted,
having an exceptionally large column of intervening neutral gas compared to
other sources we have studied. Nonetheless, the occulting material is not
manifestly dark in a local sense. About 6\mag{} of visual extinction comes
from low density (1\pccm{}) gas along the 4-5\kpc{} length of the
sightline. Comparable intervening column density of H-nuclei comes from 5-6
diffuse clouds ($\Av \about 1\mag$ each), which display a surprisingly
complex chemistry relative to their modest inferred column density.

In an attempt to characterize more fully the nature of the host gas clouds,
we undertook to map CO emission around the NRAO150 sightline.  However,
targeting NRAO150 involves making one serious tradeoff: in return for
access to kinematic complexity along with the ability to map several
strongly-absorbing clouds simultaneously and the chance to address various
puzzles raised by our previous observations, we forfeit precise knowledge
of the distances of the intervening clouds.  Only an angular scale may
readily be attached to our present conclusions, a fact typical of
observations in the galactic plane.  For reference, we note that the line
of sight velocity gradient in the disk, assuming a flat rotation curve with
$R_0 = 8.5\kpc$ and a rotation velocity of $220\kms$ implies 1) a line of
sight distance of 2.5\kpc{} for a velocity of $-20\kms$ (\ie{} at the blue
edge of our strong absorption features) and 2) a distance uncertainty of
$\pm 0.75\kpc$ associated with the typical cloud-cloud velocity dispersion
of 6\kms.  At a distance of 1\kpc{}, 1\arcmin{} corresponds to 0.3\pc{}.

\section{Observations and data reduction}
\label{sec:obs}

\subsection{Observations}

We used the IRAM Plateau de Bure Interferometer (\PdBI{}) to achieve higher
spatial resolution in mapping CO emission from the gas toward NRAO150.
However, an interferometer filters out low spatial frequencies of the
source distribution and to supply the missing but much-needed information
we observed the same region of the sky with the \IRAMthm{} telescope.  In
this Section we describe in detail how these observations were taken,
reduced and merged to form the on-the-fly (OTF) and hybrid synthesis
results.

\subsubsection{IRAM Plateau de Bure Interferometer observations}

\PdBI{} observations dedicated to this project were carried out with 5
antennae in the CD configuration (baseline lengths from 24\m{} to 229\m{})
from July to December 2005. Three correlator bands of 20 MHz were
concatenated around the \twCO{}~\Jone{} frequency to cover a $\about50\kms$
bandwidth at a resolution of $\about0.1\kms$. Two additional correlator
bands of 320 MHz were used to measure the 2.6\mm{} continuum over the
580\MHz{} instantaneous IF-bandwidth available with this generation of
receivers.

We observed a seven--field mosaic. The mosaic was centered on the position
of NRAO150 and the field positions followed a compact hexagonal pattern to
ensure Nyquist sampling in all directions. This mosaic was observed for
about 24~hours of \emph{telescope} time with 4 to 6 antennas. Taking into
account the time for calibration and after filtering out data with rms
tracking error larger than $1''$ (computed over the scan length), this
translated into \emph{on--source} integration time of useful data of
3.3~hours in the D configuration and 7~hours in the C configuration for a
full 6-antenna array.  Typical 2.6\mm{} resolution is $5.8''$ in the D
configuration and $2.6''$ in the C configuration.

\subsubsection{IRAM-30m singledish observations}

\TabFlux{} %

The \twCO{}~\Jone{} and \Jtwo{} lines were observed simultaneously during
about 4 hours of good summer weather ($\about 4\mm$ of water vapor) in
September 2005. The four available single-side band mixers were used (two
per frequencies: AB100 and AB230) in the OTF observing mode to map a
field-of-view of $220''\times 220''$ centered on the position of NRAO150.
The dump time was 1 second and the scanning speed $\about 3''/$sec.  The
field-of-view was covered twice by raster-scanning along the RA axis and
once by scanning along the Declination axis. The HPBW of the telescope for
the J=1-0 line is 22\arcsec\ and the rasters were separated by $10''$,
providing Nyquist sampling at 2.6\mm{} but undersampling at 1.3\mm{}. We
estimate the 30m position accuracy to be $\about 2''$.

We used the ON-OFF switching mode and separately observed the reference
position at $(+400'',0'')$ during 62 minutes using frequency switching mode
with a frequency throw of 7.7\MHz{} (optimal to suppress standing waves) to
check for the presence of signal.  Emission was detected at
$\lambda$2.6\mm{} and later added back into the on-the-fly emission
profiles.

\subsection{Data reduction}

The data processing was done with the \GILDAS{}\footnote{See
  \texttt{http://www.iram.fr/IRAMFR/GILDAS} for more information about the
  \GILDAS{} softwares.} software suite~\citep{pety05}.

\subsubsection{Singledish calibration}

The \IRAMthm{} data were processed inside \GILDAS{}/\CLASS{} software. They
were first calibrated to the \Tas{} scale using the chopper wheel
method~\citep{penzias73}, and finally converted to main beam temperatures
(\Tmb{}) using the forward and main beam efficiencies (\Feff{} \& \Beff{})
displayed in Table~\ref{tab:obs}.  The resulting amplitude accuracy is
\about{} 10\%. The line window was fixed between $-25$ and $10\kms$ and a
second order baseline was subtracted from each on-the-fly spectrum.  The
off-position frequency-switched spectrum was fitted by a combination of
Gaussians after baseline subtraction and this fit was then added to every
on-the-fly spectrum.

To compensate for the presence of absorption in the OTF-data near the
continuum background source, the absorption spectra measured several years
ago at \PdBI{} \citep{LisLuc98} were scaled to the flux of NRAO150 at the
observation date (Table~\ref{tab:fluxes} here), converted to main beam
temperature, convolved by a Gaussian whose full width at half maximum
corresponds to the natural resolution of the 30m at the observing
frequency, and (at last) subtracted from each on-the-fly spectrum. The
resulting spectra were finally gridded through convolution by a Gaussian.
The \twCO{}~\Jtwo{} data were gridded at the same spatial resolution (\eg{}
$22.5''$) as the \twCO{}~\Jone{} data on account of the observational
undersampling at the higher frequency.

\subsubsection{Inteferometric calibration}

The first steps of the standard calibration methods implemented inside the
\GILDAS{}/\CLIC{} software were used for the \PdBI{} data.  The
radio-frequency bandpass was calibrated using observations of 3C354.4
leading to an excellent bandpass accuracy because its 2.6\mm{} flux was
larger than 20 Jy at the epochs of the observations. The flux of NRAO150
was determined against the primary flux calibrator used at \PdBI{}, \ie{}
MWC349. The resulting flux accuracy is \about{} 15\%.

We then took advantage of the presence of the strong NRAO150 point source
continuum (Table~\ref{tab:fluxes}) to get a very accurate phase and
amplitude calibration.  $uv$ tables for the 2.6\mm{} continuum and the
\twCO{}~\Jone{} were created and time dependencies of the phase and
amplitude of the continuum were self-calibrated per baseline through the
use of the \GILDAS{}/\texttt{UV\_GAIN} task.  The temporal phase and
amplitude gain derived during this self-calibration step were applied to
the \twCO{}~\Jone{} $uv$-tables through the use of the
\GILDAS{}/\texttt{UV\_CAL} task. Those last two steps were done for each
individual mosaic field taking into account that the flux of NRAO150 is
attenuated differently by the interferometer primary beam depending on the
field position. They were also done on each individual observation session
to take into account the temporal variations of the NRAO150 flux.  Finally,
the \twCO{}~\Jone{} absorption spectrum was added back to the data after
rescaling to the flux of NRAO150 at the epoch of observation.

\subsubsection{Short-spacing processing, imaging and deconvolution}

Following~\citet{gueth96}, the \GILDAS{}/\MAPPING{} software and the
singledish map from the \IRAMthm{} were used to recreate the short-spacing
visibilities not sampled at the Plateau de Bure.  These were then merged
with the interferometric observations.  Each mosaic field was then imaged
and a dirty mosaic was built combining those fields in the following
optimal way in terms of signal--to--noise ratio~\citep{gueth01}
\begin{displaymath}
\displaystyle %
    J(\alpha,\delta) = \sum\nolimits_i \frac{B_i(\alpha,\delta)}{\sigma_i^2}\,F_i(\alpha,\delta)
    \left/
      \displaystyle \sum\nolimits_i \frac{B_i(\alpha,\delta)^2}{\sigma_i^2}.
    \right.  
\end{displaymath}
In this equation, $J(\alpha,\delta)$ is the brightness distribution in the
dirty mosaic image, $B_i$ are the response functions of the primary antenna
beams, $F_i$ are the brightness distributions of the individual dirty maps
and $\sigma_i$ are the corresponding noise values. As may be seen in this
equation, the dirty intensity distribution is corrected for primary beam
attenuation, which make the noise level spatially inhomogeneous.  In
particular, noise strongly increases near the edges of the field of view.
To limit this effect, both the primary beams used in the above formula and
the resulting dirty mosaics are truncated. The standard level of truncation
is set at 20\% of the maximum in \MAPPING{}. The dirty image was
deconvolved using the standard H\"ogbom CLEAN algorithm. The resulting data
cube was then scaled from Jy/beam to \Tmb{} temperature scale using the
synthesized beam size (see Table~\ref{tab:obs}).

Although we also obtained data in the C configuration of \PdBI{}, we
present here only the combination of the more compact D configuration with
the \IRAMthm{} singledish data. Indeed, we were not able to correctly
deconvolve the combination of C + D configuration and singledish data. We
guess that this is due to the too low signal-to-noise ratio of the C
configuration data.

\subsection{Field of view and finally-achieved resolution}

The resolution of the shown OTF singledish maps is 22.5\arcsec\ both for
the \twCO\ J=1-0 and J=2-1 lines. As noted in Table 1, the spatial
resolution of the hybrid synthesis results shown here (using the D-array
data) is 5.8\arcsec\ ($6.0'' \times 5.5''$).

The 220\arcsec\ extent and 90\arcsec\ diameter of the OTF and hybrid maps
correspond in angle to the HPBW of typically-illuminated telescopes (HPBW =
$1.2\lambda$/D) of diameter D = 2.9m and 7.2m in the J=1-0 line and the
5.8\arcsec\ synthesis beam corresponds to D = 109m.

\subsection{Relative intensities}

\FigProfileEvolution{} %

The validity of the discussion depends on the consistency of the intensity
calibration across several instruments and over a wide variety of spatial
scales.  For the \TR$^*$ scale employed at the 12m telescope, the antenna
temperature is standardly scaled up by 1/0.85 representing the fraction of
the forward response subtended by the Moon. To put the data on its standard
main-beam scale \Tmb, \IRAMthm{} antenna temperatures are scaled up by
1/0.75 (see Table 1 here) representing the fraction of the forward response
in the main beam.

To make the point that these intensity scales are consistent,
Fig.~\ref{fig:B0355f1} and~\ref{fig:ProfileEvolution} compare the profiles
toward NRAO150 at the same angular resolution from different instruments:
1) 22.5\arcsec\ spectra from the \IRAMthm{} and the smoothed synthesis data
and 2) 60\arcsec{} spectra from the Kitt Peak 12m and the smoothed
synthesis data. The thin line (red) spectrum in the upper part of
Fig.~\ref{fig:ProfileEvolution} displays the convolution of the hybrid
synthesis data to the resolution of the \IRAMthm{} telescope. The agreement
with the \IRAMthm{} spectrum shown in the top panel of
Fig.~\ref{fig:B0355f1} demonstrates the reliability of the algorithm used
to merge the singledish and interferometric data. At bottom in
Fig.~\ref{fig:ProfileEvolution}, we also show the result of convolving the
hybrid synthesis data to the 60\arcsec\ resolution of the 12m telescope
without any further amplitude rescaling. However coincindentally, the
convolved hybrid synthesis data reproduce the 12m telescope integrated
intensity to within 4 percent in either strong feature as well as the
detailed profiles themselves (although there is a very slight velocity
shift in the gas around $-10\kms$).  Therefore we are confident that the
very strong lines seen here at high resolution, 12-13\K{}, are not
artifacts of instrumental calibration.

Finally, the 22.5\arcsec\ resolution emission profile from the \IRAMthm{}
in the top panel of Fig.~\ref{fig:B0355f1} is more than twice as strong as
that from the 12m (not just 0.85/0.75) and that from the hybrid synthesis
at the top of Fig.~\ref{fig:ProfileEvolution} is 30\% stronger still.  This
profile evolution arises from increased spatial resolution.

\section{New Observational Results}

\FigAvgMoments{} %

\subsection{Emission spectra toward NRAO150}

The upper part of the top panel of Fig.~\ref{fig:B0355f1} compares the
22.5\arcsec\ resolution \IRAMthm{} spectrum from the present work with its
1\arcmin\ KP counterpart toward the continuum source. To the extent that
the $-17.5$ and $-8.5\kms$ emission features resemble themselves as the
linear resolution increases by a factor of 3, it might be inferred that the
CO emission shows little structure on sub-arcminute scales.  Conversely,
the brightness increases by almost a factor 2 at $-10.5\kms$, indicating
stronger sub-structure.

While the 22.5\arcsec\ resolution \IRAMthm{} spectrum resolves some but not
all of the disparity between the 1\arcmin\ KP CO emission profile and the
distinct doubling of the absorption components at $-10.5$ and $-8.5\kms$,
the 6\arcsec{} on-source hybrid synthesis profile in
Fig.~\ref{fig:ProfileEvolution} much more strongly resembles the absorption
profile around $-10\kms$.  So, after much work, we are finally able to say
that the disparity between the 1\arcmin\ KP12m profile and the absorption
is due to spatial gradients and the coincidental positioning of the
continuum source with respect to the foreground gas.

\subsection{Summary maps of emission integrated over each absorption feature}

\FigProfileBrightSpot{} %

To understand the overall behavior, Fig.~\ref{fig:AverageMoments} shows
summary maps integrated over five velocity ranges corresponding to the
features seen in absorption.

The feature at $-17.5\kms$ is present over the full extent of the region
mapped and/or synthesized but with substantial intensity contrast, of order
4:1 even in the middle, top panel of Fig.~\ref{fig:AverageMoments}.  At
6\arcsec{} resolution at right, there is emission toward and around the
continuum and a somewhat higher contrast, from a low of 2.1\Kkms{} (just
north of NRAO150) to 14.5\Kkms{} near the southwest rim of the synthesized
region.

The weak emission features at $-14$ and near $-4\kms$ occupy relatively
small fractions at the edges of the area mapped at 5.8\arcsec\ resolution
in Fig.~\ref{fig:AverageMoments} and approach but do not extend across the
continuum position. Stronger emission exists within the \IRAMthm{} map
outside the hybrid synthesis field of view but the beams employed here are
mostly unfilled when pointed on-source: there are no isolated hot spots
within the synthesized region which were lost to beam dilution in the
singledish data. We can thus deduce that any difficulty in detecting
emission around NRAO150 in these features is purely an excitation effect,
because the optical depths measured from the absorption profiles are of
order 0.5 in either case.

As suggested by the comparison between the 30m and 12m profiles in
Fig.~\ref{fig:B0355f1}, the sharpest spatial structure near the continuum
source occurs in the panels centered at $-10.5\kms$ in
Fig.~\ref{fig:AverageMoments}.  The ridge of emission which passes near
NRAO150 appears to have been at most barely resolved by the 30m telescope
at $-10.5\kms$, while a very similar structure at $-8.5\kms$ in the OTF
maps is more spatially extended and better resolved.

Perhaps the clearest result in Fig.~\ref{fig:AverageMoments} is the
transverse resolution and isolation of the ridge at $-10.5\kms$ near
NRAO150 in the hybrid synthesis data at right, and, to a slightly lesser
extent, at $-8.5\kms$ as well. Over the region of the synthesis the
integrated intensity contrast is high, above 16:1, but the emission does
not break up into a patchwork of unresolved and/or isolated bright and dark
regions. As discussed below, uniformity is to some extent an artifact of
the high optical depth in the middle of the ridge, enhanced by integration
over velocity which suppresses spatial structure even at the northern edge
of the ridge where (see Sect 4.5 and Fig.~\ref{fig:Kine}) it is heavily
kinematically structured.

Taken together, similarity in the third and fourth row of
Fig.~\ref{fig:AverageMoments} strongly suggests that the features seen at
$-10.5$ and $-8.5\kms$ are actually part of the same structure, no matter
that they are distinct in absorption, separated in emission at sufficiently
high spatial resolution and substantially different in the richness of
their chemistry.

\subsection{Channel maps at 1 km s$^{-1}${} resolution}

Fig.~\ref{fig:30mMoments} and~\ref{fig:HybrMoments} (available on-line)
show a series of channel maps with a resolution of $1\kms$ for the
\IRAMthm{} and \PdBI{} data. The weakly-emitting gas at $-14$, $-4.5$
and/or $-3.5\kms$ is only marginally seen near the field center and an
East-West velocity gradient across NRAO150 is resolved in the $-17.5\kms$
gas with more positive velocities seen to the South and East.  The blue
wing of emission at $-11\kms$ is disposed on either side of the source
(somewhat filling in the gaps of the distribution at $-10\kms$).

Increasing the velocity discrimination to this extent still does not show
substantially smaller scale spatial structure in the emission pattern with
the possible exception of the panel at $-8\kms$, in the near wing of the
strong emission feature.  The high optical depths measured in the
$-17.5\kms$ feature and in the components at $-10.5$ and $-8.5\kms$ may
impede viewing such structure: No matter how sharp the actual internal
structure, it can only be seen if the line of sight penetrates deeply
enough to probe it.  This behavior is illustrated in greatest detail in
Fig.~\ref{fig:Kine} and discussed in Sect.~4.5.
 
\subsection{Bright spots, area filling factors and spatial statistics}

\FigFractionalArea{} %
\FigExcessRMS{} %

Fig.~\ref{fig:ProfileBrightSpot} displays profiles selected from hybrid
maps of peak brightness in the gas around $-10\kms$ (see the top panel) and
at $-17\kms$.  There are some very strong lines in the region of the
synthesis, \Tmb = 12-13\K{}, which was hardly expected on the basis of the
3\K{} lines in the emission spectra at 1\arcmin\ in Fig.~\ref{fig:B0355f1}.
As indicated by the spatial offsets labelling the spectra in the top panel
of Fig.~\ref{fig:ProfileBrightSpot}, such unusually strong lines exist very
close to NRAO150, \ie{} just outside the on-source synthesized beam but
still inside the on-source 30m beam.  These lines imply CO J=1-0 excitation
temperatures of at least 16-17\K{}, far stronger than is typical of very
dark clouds where $\TK< 10\K$. Perhaps an analogous surprise was that of
\cite{Hei04}, whereby a very weak (0.2\K{}) \twCO\ J=1-0 line inadvertently
seen at the 30m was subsequently found in high-resolution PdBI maps to
consist of a much brighter, unresolved spicule.

The beam efficiency needed to convert observed antenna temperatures to
actual sky brightness has two aspects; the extent of the sky distribution
-- which antenna lobes are subtended -- and the degree of uniformity or
area filling factor within that extent.  In fact there is a high degree of
heterogeneity on all the scales which are resolved in the present dataset.
Fig.~\ref{fig:PixelHistograms} expresses this fact quantitatively by
showing the fractional area occupied by emission at all observed levels of
brightness.  We defined a circle of radius 45\arcsec\ within the hybrid
synthesis map and constructed histograms of the brightness distribution for
all included pixels over various channel ranges.  At left in
Fig.~\ref{fig:PixelHistograms} are the differential and cumulative
distributions for the near wings and cores of the $-10.5$ and $-8.5\kms$
features and at right are those for the $-17.5\kms$ feature.

Cartoons inset in each panel show the profile intervals which were
employed, superposed on the mean profile formed over the region.  These
mean spectra have brightnesses of 3.6 - 4.2\K{} in the line cores and
approximately 1\K{} in the selected regions of the line wings.  Noise
statistics are not necessarily the same for channels with and without
emission (after deconvolution) but for reference we constructed similar
histograms for channels in emission-free baseline regions of the spectrum
(as shown).  90\% of the noise occurs below 0.7\K{} and 95\% below 1\K{};
the nominal single-channel rms noise averaged over the region of the
synthesis is 0.47\K{} in emission-free baseline regions of the spectrum.

\FigKine{} %

Pixel histograms for the line cores are strongly populated up to 7-8\K{}
and have tails extending to 10-13\K{}.  Even pixels in the 1\K{} line wings
have higher-intensity tails extending up to 6-7\K{}, reflecting substantial
local excursions of the line centroids (see
Fig.~\ref{fig:ProfileBrightSpot} and~\ref{fig:Kine}).  Such wide
distribution of intensity ensures that fluctuations are a significant
fraction of the mean. A plot of the true spatial rms variation in each
channel, \ie{} the rms variation above and beyond that attributable to
radiometer noise (\eg{} 0.47\K{} averaged over the field of the the hybrid
synthesis and 0.33\K{} in the 30m OTF data), calculated as the square root
of the quadrature difference between the total and noise rms, is shown in
Fig.~\ref{fig:ExcessRMS}.

For the larger $220'' \times 220''$ region of the 30m OTF data this excess
rms is closely equal to the mean across the entire spectrum, except between
$-9$ and $-10\kms$ where the mean exceeds the rms by 20-30\%.  For this
gas, increasing the spatial resolution to 5.8\arcsec\ yields only a small
increase in the ratio of mean and excess rms brightness.  By contrast, the
character of the comparison changes drastically for the $-17.5$ feature at
higher resolution, where the excess rms changes from 100\% to only 50\% of
the mean at the line core when going from the \IRAMthm{} to the synthesis
data.

\FigCOcorrelation{} %

The actual pixel statistics of the brightness distributions are given in
Fig.~\ref{fig:ExcessRMS} but fluctuations equal to the mean are most
commonly recognized as characteristic of Poisson statistics, which would be
observed from a macroscopically transparent ensemble of statistically
independent clumps.  Conversely, large ratios of mean to rms brightness
might imply that the innate structure has been completely spatially
resolved.  However the situation is both more complex and more interesting,
because the synthesis data also resolves the kinematic substructure of the
features in ways which are not reflected in this discussion.

\subsection{A detailed view of line kinematics}

The full extent of the variation in CO emission can only be revealed by
examining the data at the highest available spectral resolution.
Fig.~\ref{fig:Kine} explores the kinematics on a series of tracks along and
across the rather linear structure seen around $-10\kms$. The tracks, shown
overlaid on an integrated intensity map at top left in the Figure, are
separated vertically (in Declination) by 4 pixels or 6\arcsec\ which is
very nearly one synthesized HPBW.  Spectra along the transverse track p are
shown at bottom left, where the vertical scale is elongated compared to the
others at right in the figure.  At top right is a larger-scale diagram
along track d using 30m data and below it are shown the hybrid synthesis
data along tracks a-f. Comparing the track at the top right in
Fig.~\ref{fig:Kine} with those below is a striking deconstruction of the
singledish lower-resolution data into its interferometric high-resolution
components.

The simple, essentially featureless East-West velocity gradient in the
$-17.5\kms$ gas at top right in Fig.~\ref{fig:Kine} is manifest along
strips a-e but the manner in which the velocity change occurs is striking
at high resolution, with abrupt reversals and line-splittings; it is
zig-zag and meandering rather than smooth.  The same is true for the
emission around -10\kms, in those strips (a-e) mostly above the mid-line of
the long trunk.  Along track f, the line profile over the complex is filled
in by stronger and presumably very opaque emission seen in the profile
labelled (-7.5\arcsec,-13.5\arcsec) in Fig.~\ref{fig:ProfileBrightSpot}. We
hypothesize that the line of sight penetrates deeply into the medium around
$-10\kms$ at the edge of the trunk along tracks a-d, and that kinematic
details largely disappear along the tracks e-f as the medium becomes fully
opaque in a macroscopic sense; note the excursions in the line wing at
$v<-11\kms$ even along tracks e-f. The combination of beam-smearing and the
strength of emission just south of NRAO150 combine in on-source singledish
data to obscure the double-lined structure apparent in absorption or at
high spatial resolution.

\subsection{Rotational excitation}

Insight into the nature of the host gas is hindered by the unavailability
of high-resolution observations other than of \twCO\ and the lack of high
spatial resolution even in the J=2-1 line.  However, in diffuse gas where
the CO/\HH\ ratio is small and sharply variable, where the \twCO/\thCO\ 
ratio can be strongly altered by fractionation, and where the rotational
excitation is sub-thermal
\citep{LisLuc98,SonWel+07,BurFra+07,SheRog+07,Lis07CO}, many of the
underpinnings of typical analysis of CO profiles must be questioned in any
case.

Abundance effects can be factored out somewhat by studying \twCO\ alone, as
shown in Fig.~\ref{fig:12co21-vs-12co10}.  Such plots comparing the J=2-1
and J=1-0 intensities confirm what is already evident in
Fig.~\ref{fig:AverageMoments}, that the two distributions are remarkably
similar, especially given the sub-thermal excitation in diffuse gas.  The
data indicate a nearly universal ratio of intensities, approximately
0.725:1 with relatively small differences among the various features.  The
$-14\kms$ feature, whose excitation is too weak to produce detectable
emission toward NRAO150 has the same slope in the diagram as much more
strongly emitting features, and only the gas at $-4\kms$ seems to have a
notably different, smaller slope.  A near-constant line brightness ratio,
with a slightly smaller slope, was noted previously in low column density
gas by \cite{FalPan+98}, see their Fig.~11 and references. The easiest way
to explain this ratio is to vary the degree of beam-filling of optically
thick LTE emission at $\TK = 14\K$. We however explain in the next section
why this scenario is unlikely here.

\section{Modelling}

\subsection{Density}

To get some feeling for the observed properties of the CO emission and the
density in the host gas, we performed non-LTE microturbulent radiative
transfer calculations in stationnary models of diffuse cloud.  These models
self-consistently determine the kinetic temperature, \HH\ and CO abundances
and CO excitation temperature in uniform density ($n(\mbox{H})= n(\mbox{H
  I}) + 2 n(\HH)$) gas spheres. The column density thus varies across the
sphere, implying that the modelled profiles samples very different column
densities, excitation conditions and chemical behaviors. These models are
described in details in \cite{LisLuc00} and \cite{Lis07}. They were used
by~\citet{Lis07CO} to interpret optical observations of CO excitation in
diffuse clouds \citep{LisLuc98, SonWel+07,BurFra+07,SheRog+07}.

\FigModelTBvsTB{} %

Results for the models are given in Fig.~\ref{fig:ModelTBvsTB} and they
show that the observed line ratios are as expected for non-LTE CO emission
from cool media of low-moderate number and column density $n(\mbox{H}) = 64
- 512 \pccm$, $N(\mbox{H}) < 3.8\times10^{21}\pscm$ or $\AV < 2\mag$.  The
pressures in these models are $p/k = (2-8) \times 10^3\Kpccm$ and the \HH/H
ratio varies considerably (increasing with $n(\mbox{H})$ or with
$N(\mbox{CO})$, hence \TB\ and the other results are line-of sight averages
like the data).  The observational noise levels are sufficient to obscure
the small differences in line ratio which would distinguish between model
densities for J=1-0 brightnesses below 2\K{} and useful differences between
the model results at various total densities are really only apparent in
somewhat stronger lines.

The gas around $-4\kms$ is distinguished in the rightmost panel of
Fig.~\ref{fig:12co21-vs-12co10} by its locus below the indicated line of
slope 0.725 and this deviation is readily interpreted in terms of the model
results for somewhat lower density perhaps $n(\mbox{H}) = 64-128 \pccm$,
although small line ratios also arise in the outer regions of model spheres
with higher density if the column density is small enough.  By contrast,
the loci for the gas at $-17.5\kms$ and $-10.5\kms$ are slightly above the
line at intermediate brightness, arched like the model results, suggesting
density as high as $n(\mbox{H}) \ga 256 \pccm$.  Even these are very low
densities by the standards of molecular emission-line work, but they are at
the high end of the range for diffuse clouds with a substantial residual
fraction of atomic hydrogen.  Somewhat lower densities would have been
inferred if the collision rates of \cite{BalYan+02} for CO excitation by
H-atoms were employed, but these seem now to have been discredited
\citep{SheYan+07}.

\subsection{Bright peaks in diffuse gas}

In emission, any single-dish profile can only be regarded as a trade off
between the brightness and the beam filling factor, only the product being
actually observed. On the other hand, in absorption, once the apparent
optical depth is large, the filling factor must approach unity.  Because
the $-17.5$ and $-10.5\kms$ features have large apparent optical depths in
absorption toward NRAO150 where they are somewhat weaker in emission, they
must occult the continuum fully (at $-17.5\kms$) or very nearly so
($-10.5\kms$). In this case, it is likely that the synthesized beam area is
filled, because, if the medium were very patchy there would be a high
chance that a background source could escape occultation and absorption
even when a foreground cloud was visible in CO emission.  Because this does
not occur in our work we infer that the beams used to probe the CO are
filled or very nearly filled by the optically thick emission, leaving
little room for an increase in the beam filling factor.
  
Assuming that the beam is filled, the models illustrated in
Fig.~\ref{fig:ModelTBvsTB} shows that 10\K{} and brighter CO J=1-0 lines
may be produced in diffuse clouds.  These occur for $n(\mbox{H}) \ga
256\pccm$, $N(\HH) \ga 10^{21}\pscm$ $N(\mbox{CO}) \ga 10^{16}\pscm$ when
the free gas phase carbon is on the verge of recombining from C\p\ to CO
and the fractional abundance $X$(CO) varies very rapidly with either the
number or column density of hydrogen, see Fig.~1 and Fig.~6 of
\cite{Lis07CO}.  Even though the CO lines are already optically thick,
observations of CO in absorption in diffuse gas show that the integrated
brightness of the J = 1-0 line is proportional to the CO column density
({\it ibid}). This observational fact is indeed typical of subthermal
excitation as predicted long ago by~\citet{goldreich74}. Hence, a small
increase in number or column density of hydrogen at the above threshold may
greatly increase the CO column density, which in turn produces brighter
optically thick CO lines.
  
Bright CO lines around NRAO150 thus most probably reflect the variation of
a chemical process, \ie{} the C\p-CO conversion. However, the ultimate
cause of the variations of this chemical process in such a limited field of
view remains uncertain.

\section{Summary; prospects for future work}

After a long series of investigations into the absorption line chemistry of
galactic diffuse gas coincidentally seen toward distant radio point sources
at low-moderate galactic latitude, we undertook to image the absorbing
medium by mapping \twCO\ in emission around one source at the highest
practicably obtainable spatial resolution.  There are many questions which
might be asked of such data and the comparisons between relatively
microscopic (sub-arcsecond pencil absorption beams and the much larger
($6-60''$) fan beams in emission.  The present work is a small step in this
direction.

Some of the results are rather direct, for instance concerning the various
disparities and differences between the absorption profiles and the
existing 60\arcsec-resolution KP CO data (Fig.~\ref{fig:B0355f1}).  An
atypical disparity between the strong emission and absorption profiles at
$-10 \kms$ was shown to arise because of insufficient spatial resolution in
the emission and not, for instance, due to the presence of a warmer (more
highly-excited) CO component seen in emission with low optical depth.  The
weakness of the $-14$ and $-4.5/-3.5\kms$ emission lines toward NRAO150,
which are seen in absorption at moderate optical depth (0.5) in
Fig.~\ref{fig:B0355f1}, arises because the regions of sufficient
collisional excitation and stronger emission are somewhat removed from the
continuum position.  These clouds, which are common in our previous
absorption surveys, are not a previously-hidden large scale molecular
component revealed in absorption, but they do highlight the robustness of
the chemistry into regions where the density and column density are too low
to produce much rotational excitation, even in CO.  This is in fact the
norm in optical absorption studies of CO.

We explored the spatial structure of the gas in several ways, \ie{} by
examining the maps at 22.5\arcsec\ and 6\arcsec\ resolution
(Fig.~\ref{fig:AverageMoments}) and by calculating the spatial statistics
both as profile rms over the mean velocity profile
(Fig.~\ref{fig:ExcessRMS}) and as channel by channel probability
distributions over the synthesized region (Fig.~\ref{fig:PixelHistograms}).
At the lower resolution, the velocity profile of rms fluctuations due to
spatial structure, calculated as the channel by channel quadrature
difference between the total rms and that seen in line-free channels,
closely resembles the mean profile, except in a narrow interval near the
core of the $-10\kms$ line.  Such statistics are quasi-Poisson in
character, consistent with a random macroscopically transparent ensemble of
independent clumps.  However, when the gas is studied at higher resolution,
where there is an even wider range of observed brightness at all
velocities, the statistics in the core of the $-17.5\kms$ noticeably change
character and the rms is only 50\% of the mean (Fig.~\ref{fig:ExcessRMS}).

In this work, it appears that every profile seen at 22.5\arcsec\ with a
linewidth greater than that attributable to the local sound speed
(full-width at half-maximum $= 0.96\kms$ for a pure-\HH\ gas with $\TK =
40\K$) is found to be composite when viewed at higher resolution:
decomposable into multiple narrower components and/or having
spatially-resolved velocity gradients in narrower-lined gas.  Whether this
would continue in the broadest components seen at 5.8\arcsec\ resolution is
an open question.

The full extent of the emission structure is realized only when the highest
spatial and velocity resolution are viewed together in position--velocity
tracks along portions of the gas where the medium may be supposed to be
macroscopically optically thin (Fig.~\ref{fig:Kine}).  A series of tracks
parallel to but somewhat displaced from the peak of a particularly well
defined filamentary or trunk structure in the gas around $-10\kms$ was used
to show the substructure in what are almost featureless position--velocity
diagrams at 22.5\arcsec\ resolution.  The high resolution version resolves
a monotonic and rather bland velocity gradient in the $-17.5\kms$ feature
into a striking series of kinks and abrupt reversals which strongly
resemble simulations of turbulent flow of the sort shown by
\cite{FalLis+94}.

The flat profile core of the $-10\kms$ component seen at 60\arcsec\ 
resolution, which is slightly split at 22.5\arcsec\ resolution
(Fig.~\ref{fig:B0355f1} and Fig.~\ref{fig:ProfileEvolution}), shows an
underlying triple nature (Fig.~\ref{fig:ProfileBrightSpot} and
Fig.~\ref{fig:Kine}) whose individual components also show this striking
pattern of velocity reversals when the medium is macroscopically thin and
show some excursions in the line wing where it is thick.  In this
connection, note that, in our data, it was not necessary to avoid the
strongly-emitting optically thick line cores (at least they are optically
thick toward NRAO150) in order to see such structure.  Furthermore, as
noted just before, when studied at lower angular resolution, regions of the
line core exhibit the same fractional level of fluctuation as the line
wings for the component at $-17.5\kms$.  At $-10\kms$ the level of
fluctuation is not very different at the two resolutions, and is still
relatively large in the line core.

In the large, the CO excitation in the host gas sampled in the J=1-0 and
J=2-1 lines at 22.5\arcsec\ resolution (Fig.~\ref{fig:12co21-vs-12co10}) is
readily explained as arising in low-moderate density n(H) $= 50 - 250
\pccm$ diffuse gas (Fig.~\ref{fig:ModelTBvsTB}).  The 12-13\K{} \twCO\ 
lines seen in Fig.~\ref{fig:ProfileBrightSpot}, at the high end of the
brightness distribution (Fig.~\ref{fig:PixelHistograms}) are probably
regions of higher density (n(H) = 300 - 500 $\pccm$) and column density
N(\HH) $> 10^{21} \pscm$, N(CO) $> 10^{16} \pscm$ where C\p-CO conversion
is more advanced even while the temperature is still elevated, $\TK >>
10\K$.  It is something of a coincidence that the \twCO\ brightnesses
observed in diffuse and dark gas are so similar, given the profound
differences in abundance and excitation: excitation in diffuse clouds is
considerably sub-thermal and the temperatures are high, typically above
30\K{}, and only a small fraction of the gas-phase carbon nuclei reside in
CO -- typically 5\% or less. This coincidence will complicate the
interpretation of CO emission in general.

While absorption studies enable very precise determination of various
chemical tracer column densitites, they can currently be done only on a
limited number of pencil beam lines of sight. Those studies thus miss
important information about the environment such as the geometry and the
kinematics of the host gas, information which could help to unravel
remaining mysteries like the abundance of \hcop{} or CH\p~\citep[see
\eg{}][]{falgarone06,hilyblant07,hilyblant08}.  This paper is the first of
a series in which we image diffuse gas around lines of sight which have
previously been studied in absorption.  In this series our observations
will range from the highest possible angular resolution in small fields of
view (as as present) to lower angular resolution in fields of view wide
enough to contain the whole structure of the host gas.  In this way we will
relate such different gas and dust tracers as H I, CO, \hcop{}, extinction,
{\it etc.} using whatever means are available to render the diffuse
absorbing material both more readily visible and more completely
understood.

\begin{acknowledgements}
  We are grateful to the IRAM staff at Plateau de Bure, Grenoble, Granada
  and Pico Veleta for their support during the observations.  The National
  Radio Astronomy Observatory is operated by Associated Universites, Inc.
  under a cooperative agreement with the US National Science Foundation.
  We are grateful to the referee for remarks which inspired improvements in
  the manuscript.
\end{acknowledgements}
 
\bibliographystyle{aa} %
\bibliography{ms9803} %

\Online{} %
\FigThmMoments{} %
\FigHybMoments{} %
\FigFractionalAreaTwo{} %
\FigMoments{} %
\FigBrightSpot{} %

\end{document}